\documentclass[manuscript]{emulateapj}

\slugcomment{Submitted to The Astrophysical Journal}

\shorttitle{Kepler Mission Noise Properties}
\shortauthors{Gilliland, et al.}
\def\Msun{\,{\rm M}_\odot}

\begin{document}


\title{KEPLER MISSION STELLAR AND INSTRUMENT NOISE PROPERTIES}


\author{
Ronald L. Gilliland,\altaffilmark{1}
William J. Chaplin,\altaffilmark{2}
Edward W. Dunham,\altaffilmark{3}
Vic S. Argabright,\altaffilmark{4}
William J. Borucki,\altaffilmark{5}
Gibor Basri,\altaffilmark{6}
Stephen T. Bryson,\altaffilmark{5}
Derek L. Buzasi,\altaffilmark{7}
Douglas A. Caldwell,\altaffilmark{8}
Yvonne P. Elsworth,\altaffilmark{2}
Jon M. Jenkins,\altaffilmark{8}
David G. Koch,\altaffilmark{5}
Jeffrey Kolodziejczak,\altaffilmark{9}
Andrea Miglio,\altaffilmark{2}
Jeffrey van Cleve,\altaffilmark{8}
Lucianne M. Walkowicz,\altaffilmark{5}
William F. Welsh,\altaffilmark{10}
}
\altaffiltext{1}{Space Telescope Science Institute, 3700 San Martin Drive,
Baltimore, MD 21218; \mbox{gillil@stsci.edu}}
\altaffiltext{2}{School of Physics and Astronomy, University of Birmingham, UK}
\altaffiltext{3}{Lowell Observatory, Flagstaff, AZ 86001}
\altaffiltext{4}{Ball Aerospace and Technologies Corp., 1600 Commerce St., Boulder CO  80301}
\altaffiltext{5}{NASA Ames Research Center, Moffett Field, CA 94035}
\altaffiltext{6}{Astronomy Department, University of California, CA 94720}
\altaffiltext{7}{Eureka Scientific, 2452 Delmer St., Suite 100, Oakland, CA 94602}
\altaffiltext{8}{SETI Institute/NASA Ames Research Center, Moffett Field, CA 94035}
\altaffiltext{9}{Space Science Office, NASA Marshall Space Flight Center, Huntsville, AL 35812}
\altaffiltext{10}{Astronomy Department, San Diego State University, San Diego, CA 92182}



\begin{abstract}
{\em Kepler Mission} results are rapidly contributing to fundamentally new
discoveries in both the exoplanet and asteroseismology fields.  The data
returned from {\em Kepler} are unique in terms of the number of stars
observed, precision of photometry for time series observations, and the
temporal extent of high duty cycle observations.  As the first mission to
provide extensive time series measurements on thousands of stars over
months to years at a level hitherto possible only for the Sun, the results
from {\em Kepler} will vastly increase our knowledge of stellar variability
for quiet solar-type stars.
Here we report on the stellar noise inferred on the timescale of a few
hours of most interest for detection of exoplanets via transits.
By design the data from moderately bright {\em Kepler}
stars are expected to have roughly
comparable levels of noise intrinsic to the stars and arising from a
combination of fundamental limitations such as Poisson statistics and any
instrument noise.  The noise levels attained by {\em Kepler} on-orbit
exceed by some 50\% the target levels for solar-type, quiet stars.
We provide a decomposition of observed noise for an ensemble of 
12th magnitude stars arising from fundamental terms (Poisson
and readout noise), added noise due to the instrument and that intrinsic
to the stars.  The largest factor in the modestly higher than anticipated
noise follows from intrinsic stellar noise.
We show that using stellar parameters from galactic stellar synthesis
models, and projections to stellar rotation, activity and
hence noise levels reproduces the primary intrinsic stellar noise features.
\end{abstract}


\keywords{methods: observational --- stars: oscillations --- stars: activity ---
stars: late type --- stars: statistics --- techniques: photometric}



\section{INTRODUCTION}

  As of early 2011 the {\em Kepler Mission} is roughly half way through its
baseline 3.5 year mission, with about one-third of the 
baseline data having now received some analysis.   
The pace of discoveries and results from the {\em Kepler Mission}
are accelerating and already quite extensive, even though the 
prime goal of detecting true Earth-analog exoplanets remains in 
the future, simply from needing to have more extensive time coverage
in order to detect transits spaced by a full year.
In the area of exoplanets, over 1200 candidates have been detected
and reported in \citet{bor11}, with validation of exoplanets first
following from transit timing variations discussed by \citet{hol10},
the smallest rocky planet to date in \citet{bat11}, and a six-planet
system unveiled in \citet{lis11}.
Discoveries from {\em Kepler's} second area of emphasis are 
equally impressive ranging from detection of oscillations on
solar-type stars for hundreds of cases in \citet{cha11a} (only
tens had been known before the {\em Kepler Mission}) to detailed 
inferences on individual stars (\citealp{met10,kur11}), and
a recent breakthrough involving the detection of both the expected
p-modes as well as g-modes allowing sensitive new studies of red giant
stars (\citealp{bed11,bec11}).
In stellar astrophysics
{\em Kepler} has revealed a remarkable eccentric A-star binary
viewed nearly face on with reflected light features superimposed
on tidally-forced g-mode oscillations \citep{wel11}, and two
instances of mutually eclipsing hierarchical triple systems
(\citealp{car11,der11}).

Returning to a discussion of the prime goal of detecting true
Earth-analogs, the observational quest is truly daunting.  An 
Earth-analog presents a transit signature with a depth of 85
parts per million (ppm), lasting a statistical average of 10 hours,
and occurring once per year.  The signal would only be evident after
seeing three successive transits equally spaced in time.  
Since the geometric probability of having transits in Earth-analog
systems is only 0.5\%, it is necessary to monitor a very large 
number of stars nearly continuously over multiple years at very high precision to have
any chance of success.  The {\em Kepler Mission} was scoped
\citep{koc10} to be able to determine the frequency of Earth-size
planets at sufficiently long orbital periods to reach well into
the habitable zone \citep{kas93}.  This is challenging on a number
of levels, including perhaps the simplest one of the noise levels
needed to support detection.  Notable in the \citet{bor11} paper is a significant
fall-off in planet frequency for the smallest planet candidates,
especially at the long orbital periods associated with habitability
(possible presence of liquid water on the surface based on equilibrium
temperatures).  Some of this fall-off is expected to date, since 
insufficient time has passed to be able to detect true Earth analogs.
However, the relative sparsity of Earth-sized planets at shorter 
orbital periods is open to interpretation and will surely be the
focus of much discussion in the exoplanet community.  Do the 
statistical trends primarily follow from the physics of planet formation
(\citealp{gou11,ida04}), or is there a strong contribution
from observational incompleteness given the larger than anticipated
noise levels (\citealp{jen10b,chr10})?  This question is beyond
the scope of this paper to resolve.  We will endeavor to more
fully explore the noise levels obtained in the {\em Kepler Mission}
given that these are not as low as expected.

The approach in the remainder of the paper is as follows.
A self-contained introduction to the mission 
and data characteristics will be given in \S 2.
Based on consideration of the {\em Kepler} data set itself, 
and knowledge of instrument characteristics such as detector
readout noise, a subset of $\sim$12th magnitude stars will be used
to establish a decomposition of noise into temporal, 
instrument, and intrinsic stellar terms in \S 3.
\S 4 will provide simulations of intrinsic stellar noise
as a means of assessing the reasonableness of the noise
separation, and for motivating what may be learned about stellar astrophysics from
the {\em Kepler} results.
A fainter sample of stars will be reported on in \S5.
Prospects for confirmation of
the stellar noise differences found will be touched upon
in \S 6 with a summary in \S 7.

\section{\em KEPLER \rm OBSERVATIONS AND NOISE METRIC}

  Characteristics of the {\em Kepler Mission} have been
presented in detail in \citet{koc10} with details on
the pipeline processing given in \citet{jen10a}.  The overall
instrument performance is discussed by \citet{cal10}, with target selection
covered by \citet{bat10}, operations by \citet{has10} and initial data
characteristics given by \citet{jen10b} and \citet{gil10}.
\citet{cia11} provide a general survey of stellar variability
based on the first full month of {\em Kepler} data.
Here we discuss some of the characteristics that are most pertinent
to understanding the overall noise levels, both those 
arising from fundamental limitations like Poisson statistics
on the sources, and simple instrument limitations such as CCD
readout noise.
  We consider only the Long Cadence (LC) data, which consist of
summations of 270 individual readouts each a little over
6 seconds long into 29.4 minute intervals. 
In addition we limit our attention to those intermediate
time scales that are most relevant for the detection of
transits.  At the high precision levels provided by {\em Kepler}
it has been found that all red giants are obvious photometric
variables \citep{koc10}; we will endeavor to ignore these and other classes
of variable stars by restricting our attention to roughly
solar-type stars.

  The {\em Kepler} field of view was carefully selected to
fall relatively near, but not on, the galactic plane.  This 
provides a high stellar density of dwarf stars, while avoiding
too high a density of fainter background stars, which could
lead to false positives through diluted
eclipses of background binaries.  
The primary range of stellar brightness followed by {\em Kepler}
is {\em Kp} = 9 to 15 \citep{koc10}, where {\em Kp} represents a broad bandpass
from about 420 to 900 nm roughly a combination of $V$ plus $R$ bandpasses.
Saturation occurs at about {\em Kp} = 11.5, but conservation of charge
is preserved beyond saturation and excellent photometry is easily recovered from
targets up to 7 magnitudes beyond saturation by summing over
the pixels bled into -- along columns \citep{gil10}.  For special applications
on highly variable stars, observations to at least 20th magnitude
may be useful.  The point design for {\em Kepler} is based
on results at 12th magnitude, and we will concentrate on 
stars with {\em Kp} = 11.5 to 12.5 for which fundamental noise
terms, instrument contributions, and intrinsic stellar noise
may be comparable in scale, thus facilitating unique determination
of the separate levels.

  The primary noise metric for {\em Kepler} is referred to
as CDPP, Combined Differential Photometric Precision, which is
intended to be either the observed noise in a carefully
specified temporal domain, or the predicted noise level in
the same temporal domain from rolling up all contributing 
factors including intrinsic stellar variability -- see \citet{jen10b}
for discussion of early on-orbit results.  By design CDPP
near 6.5 hours (one-half the time for a central transit of 
a true Earth-analog) was intended to be about 20 ppm for the 
mode of dwarf stars at 12th magnitude, in order to provide
slightly more than 4-$\sigma$ signal-to-noise ratio for single 
transits with 85 ppm depth.
With four such transits the overall
detection statistic \citep{jen02} would then reach a level
providing less than one statistical false positive over the
whole ensemble of such stars searched for transits.
In reality \citep{chr10} the CDPP assessed as the mode
over 12th magnitude dwarf stars is about 30 ppm, or 50\%
higher than planned for.  
Based upon considerations of solar variability in the 
relevant temporal domain \citep{jen02}, stars were allocated
10 ppm in the noise budget to be rolled up into the CDPP.
With a Poisson term due to counting some 4.5 $\times 10^9$ e-
per 6.5 hours at 12th magnitude equal to about 15 ppm, the
stellar noise was expected to be a significant contributor
to the total noise, but not usually a limiting factor.
A tail to higher intrinsic stellar noise was anticipated
from more active, typically younger stars.

Having determined the observed CDPP for a star, or ensemble
of such stars, how would one determine what part of this
is intrinsic to the star?  Answering this question will 
take up most of \S 3.  We continue here with an introduction
to the observations and data characteristics that will need
to be accounted for and analyzed fully to address this.

The {\em Kepler} data are collected in segments one-quarter
of the 372-day spacecraft-year long.  {\em Kepler} is in an 
Earth-trailing orbit which provides unobscured access to
a field well off the ecliptic throughout the year.  In order
to keep the solar arrays illuminated, successive 90 degree
rolls are executed every one-quarter year.
The focal plane consists of 42 CCDs with 27 $\mu$ pixels
(with very deep wells of about 1.1 $\times 10^6$ e-) and
1024$\times$2200 rows and columns.
Each CCD has two readout amplifiers leading to a total 
of 84 channels.  The CCDs are mounted pair-wise into
21 modules which cover a 5$\times$5 grid except for 
the corners which have smaller pixel scale CCDs used
only for precisely guiding the spacecraft.  The CCDs 
are oriented such that rotational symmetry is maintained
except for the central-most module.  Although
stars fall on a unique detector during each observing Quarter, they
maintain the same symmetry with respect to rows, columns
and stellar neighbors after each roll
(except for the central module where the stars always remain,
and symmetry is broken).

\begin{figure}
\includegraphics[width=10cm]{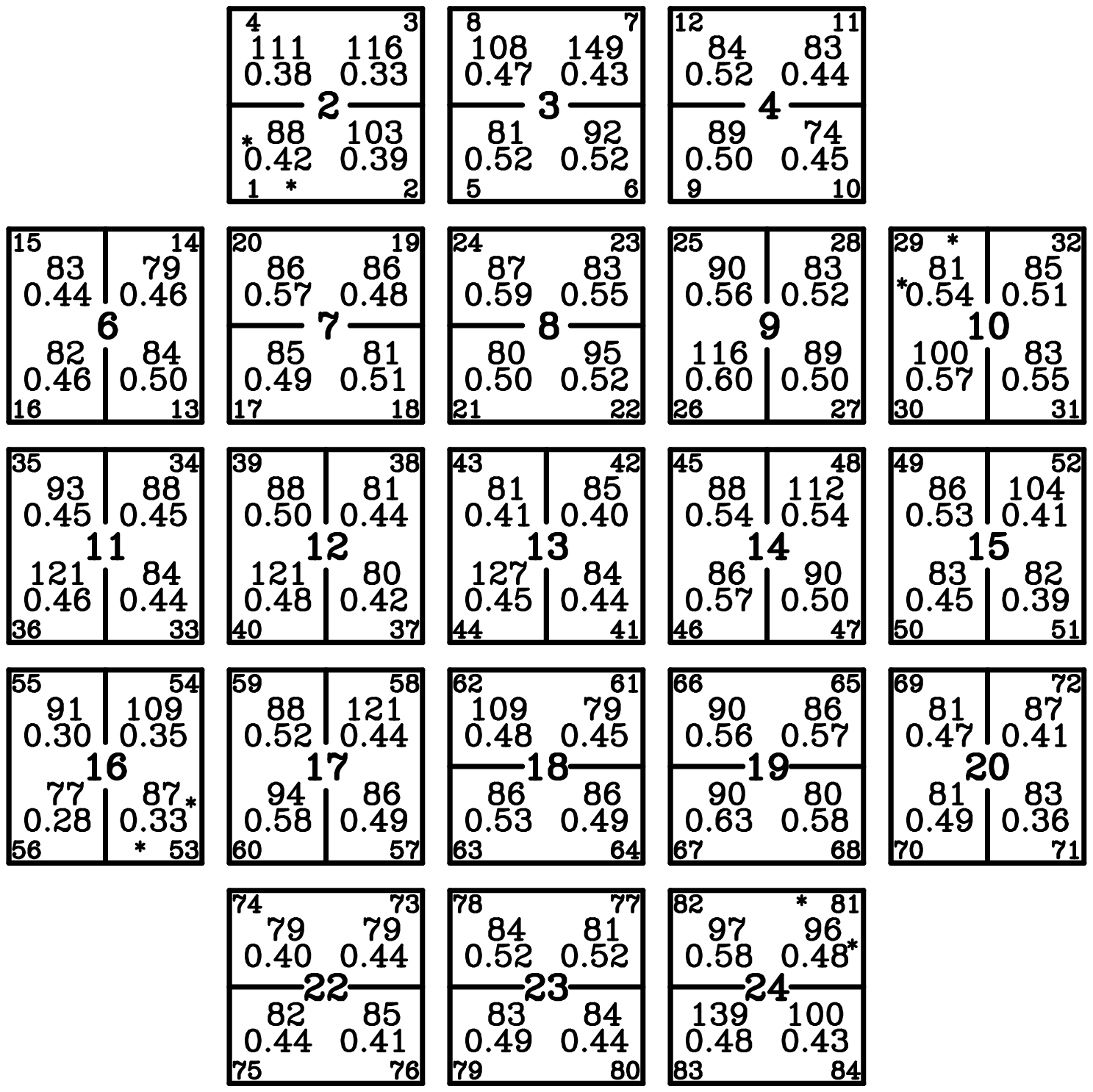}
\caption{Schematic layout of the 42 CCDs comprising the science
array for {\em Kepler}.  The CCDs are mounted pair-wise into 
21 modules labelled consecutively 1 through 25 except for the 
missing corners which have non-science CCDs used only for guiding.
Within a module each CCD is read out by two amplifiers which 
are shown on the same side of a dividing line.  The CCDs are 
laid out to maintain symmetry of rows and columns as the 
spacecraft is rotated by 90 degrees (except for central module
lacking this symmetry).  Numbers in the corners of each module
are channel numbers 1 through 84.  The pairs of numbers near
the center of each channel are (top) readout noise in electrons,
and (bottom) PSF energy content of central pixel for a centered
star.  Note that channels 1, 29, 81, and 53 are flagged with a
pair of `*', these comprise a logical quartet as discussed 
in the text.
\label{fig:gm.obs}}
\end{figure}

Fig.~\ref{fig:gm.obs} is used for several orientation 
purposes to the {\em Kepler} observing procedure and 
basic characteristics of the instrument.
The layout of CCDs illustrates the rotational symmetry
across observing Quarters.  Of greater relevance for this
paper is to note the four channels in Fig.~\ref{fig:gm.obs}
flagged as comprising one `quartet'.  For this example 
a star on channel 1 in Quarter 2 (July -- September 2009)
would be moved to channel 53 for Quarter 3 (October -- December 2009),
then channel 81 for Quarter 4 (January -- March 2010), then
channel 29 for Quarter 5 (April -- June 2010), before 
starting the cycle over on channel 1 for Quarter 6 (July -- September 2010).
The stars that started in Quarter 2 on any one of channels
1, 29, 53 and 81 will simply permute through these same channels
and thus form a natural `quartet'.  There are 21 such quartets, 
and analysis of the stars within quartets will play a key roll
in the noise decomposition discussed in the next section.

Also presented in Fig.~\ref{fig:gm.obs} are numerical values
by channel for readout noise in electrons for each amplifier
and a metric of the focus as the fraction of energy contained
in the central pixel for a star centered on a pixel.
Readout noise levels range from 77 to 149 e- and are 
taken from \citet{van09}.  The central pixel energy fractions
were derived by the 1st author from images taken during commissioning, 
after the primary mirror had been adjusted in tilt and piston to attain
best focus. 
A check with Q5 data one year later showed nearly identical focus values.

The 27 $\mu$ CCD pixels span a scale of 3{\farcs}98 on 
the sky, very large compared to the diffraction limit
of the 0.95-m Schmidt telescope.  Initial plans had
called for the telescope point spread function (PSF) to be
soft, specifying a maximum flux of some 30\% in the 
central pixel of a best focused star centered on a pixel.
Mid-way through mission development this was reconsidered.
Sharper PSFs would help both in terms of avoiding large
numbers of false positives through background eclipsing
binaries and optimally supporting through-transit centroiding
to eliminate most background eclipsing binaries as 
false positives that blend with the targets.
The resulting {\em Kepler} PSFs are very sharp -- several of the 
modules have central energy concentrations for a central
pixel in excess of 50\%, with the sharpest focus channels
reaching above 60\%.  The channels with the worst focus
still come in near 30\% central pixel energy fraction, 
with a mean over the full array being at 47\%.  By contrast
{\em HST} imaging instruments such as ACS or WFC3 have
typical central energy fractions more like 20\% -- for a
mission intended to have high resolution as a primary
attribute.  Simulations and now experience have both
indicated that the best photometry is obtained from 
the {\em Kepler} channels having the better focus.
While the {\em Kepler} PSFs are severely undersampled
(in many cases with the underlying FWHM of the optical
PSF being only one-quarter of a pixel), the spacecraft
guiding at the pixel scale is exquisite, much better
than the already impressive case of {\em HST}.  On the 
timescales most relevant for transits and CDPP the 
spacecraft jitter is about 6$\times 10^{-5}$ pixels.
The point-to-point jitter of the {\em Kepler} observations
is $\sim 1.2 \times 10^{-4}$ pixels.
For {\em HST}, point-to-point jitter on comparable timescales
is $\sim$0.1 pixels \citep{gil05} on the much finer 0{\farcs}05 
pixel scale of ACS.  In an absolute sense the {\em Kepler}
jitter bests {\em HST} by a full order of magnitude.
Translated to pixel scale, the {\em Kepler} advantage
is three orders of magnitude.
On longer timescales of months differential velocity aberration
can move images by over half a pixel in extreme cases over one Quarter.
The {\em Kepler} PSFs average about 4 pixels in diameter
for 95\% encircled energy, thus allowing apertures used
for photometric extractions to remain small.

At 12th magnitude the average stellar image will consist of 12-13
pixels forming the `optimal aperture'.  A `postage stamp' of pixels
including this optimal aperture used for the photometric extraction
plus a buffer halo of 1 -- 2 pixels radius is returned 
for each target.  This optimal aperture is
held fixed within each Quarter and is selected to maximize
the signal-to-noise of each star for simple aperture sum
photometry which is the only method used in the pipeline
\citep{bry10a}.  The {\em Kepler} CCDs have large readout
noise levels by ordinary standards, at about 100 e-.
Furthermore, each Long Cadence observation consists of 
270 individual integrations.  Therefore the total variance
from readout noise on a 12th magnitude star in the 6.5 
hour intervals for which CDPP is defined sums to
about 4.5 $\times 10^8$ e-$^2$, but this is only one-tenth
the variance (direct counts) from Poisson statistics on 
the target itself.  The number of pixels used for the 
optimal aperture of a star each Quarter is a complex function of
the PSF, and whether
a target has nearby neighbors, see \citet{bry10b}.

At the middle of the 3.5 year baseline mission
the overall spacecraft status is healthy.  The only significant
degradation to date was the loss of the two CCDs on module 3
about nine months into the mission, about one month into Quarter 4.
The associated electronics failure is not expected to affect
other modules.

We adopt a noise metric that is very similar to the formal,
wavelet-based CDPP evaluated in the {\em Kepler} Science Operations
Center pipeline \citep{jen10b}, but easier to compute and explain.
Our noise metric starts with the pipeline calibrated, or PDC
(Presearch Data Conditioned) light curves archived for the project 
at the Multimission Archive (MAST) at STScI.\footnote{
http://stdatu.stsci.edu/kepler/}
These data are then subjected to a high-pass Savitsky-Golay filtering
by fitting a quadratic polynomial 2 days (97 LC points) wide
centered on each point in the time series, then subtracting this.
Sigma-clipping is introduced at 5-$\sigma$ to eliminate highly 
deviant points.  Then the data are block-averaged into 6.5-hour
segments by forming successive averages over 13 consecutive LC intervals.
Only segments containing at least 7 valid intervals are retained.
The CDPP is simply obtained as the standard deviation of the 
6.5-hour means.  The Savitsky-Golay filter suppresses (based on direct tests with
trial time series) 16.8\% of the expected rms from white Gaussian noise
for 6.5-hour bins, this is corrected for by multiplying the resulting
CDPP by $\times$1.168 to adjust for how much white noise would be
suppressed by this band pass filtering.
We thus end up with a strongly bandpass-limited
metric, having first removed any slow variations captured by a 2-day wide least squares
quadratic fit, then averaging out all time scales shorter than 6.5 hours.
The scale of this CDPP is nearly identical to the 
12-hour CDPP provided by the SOC pipeline, and about 15\% smaller
than the 6-hour CDPP from SOC.  Agreement between the two estimates
is quite good with a 1-$\sigma$ scatter of 10\%.

Such a CDPP measure is formed for each star of interest separately
for each of Quarters 2 through 6.
Q1 was omitted because it was less than half a Quarter,
and Q0 was an even shorter commissioning period using a
non-standard target set.
Module 3 was lost about three weeks into the start of Q4;
these truncated time series are also omitted.

For our purposes of determining what components of noise 
should be ascribed to the stars themselves, or other
factors, we restrict attention to stars that are not known {\em a priori}
to be variable.  At the precisions reached by {\em Kepler} all 
red giants are variable (\citealp{koc10,gil08}).  Many stars hotter than 6500 K
fall into the classical instability strip and will be variable.
Some stars are known to have variability in their light curves
from being eclipsing binaries, or Kepler Objects of Interest hosting
planet candidates.  We start with the roughly 165,000 full list of 
Kepler stars and adopt the Kepler Input Catalog (KIC) \citep{bro11} parameters.  The 
approximately 15\% of stars in the KIC that are not ``classified", i.e.
do not have T${\rm _{eff}}$, gravities, etc., are dropped from consideration.
We evaluate the predicted ${\rm \nu_{max}}$ value
(location in a power spectrum of largest p-mode amplitudes
which in the Sun is at 3150 $\mu$Hz) of solar-type stellar
oscillations and keep only the set between 1000 -- 6000
$\mu$Hz, thus quite effectively eliminating both giants and 
subgiants from the sample (see \citealp{cha11b}).
${\rm \nu_{max}}$ is proportional to gravity modified by ${\rm T_{eff}^{-1/2}}$;
our sample has been chosen to have log g $\sim$ 4.0 -- 4.7.
We also require
that the T${\rm _{eff}}$ be between 5200 - 6500 K, and that mass be
less than 1.4 $\Msun$.  The resulting list is then cross-correlated with
the known EBs and KOIs and all of these are dropped.  Stars are further restricted to have
a maximum contamination over the multiple Quarters of less than 20\%,
i.e. at least 80\% of the light in optimal apertures is thought to
come from the target star rather than light from nearby neighbors.
With all of these eliminations in place
a list of 69,005 stars remains.

At {\em Kp} = 11.5 to 12.5 used for primary results in this paper,
very few stars have large contamination levels.  Only 1\% of 12th
magnitude stars are above the 20\% cut, and only 1\% of the
retained sample exceed a 10\% contamination.

\begin{figure}
\includegraphics[width=10cm]{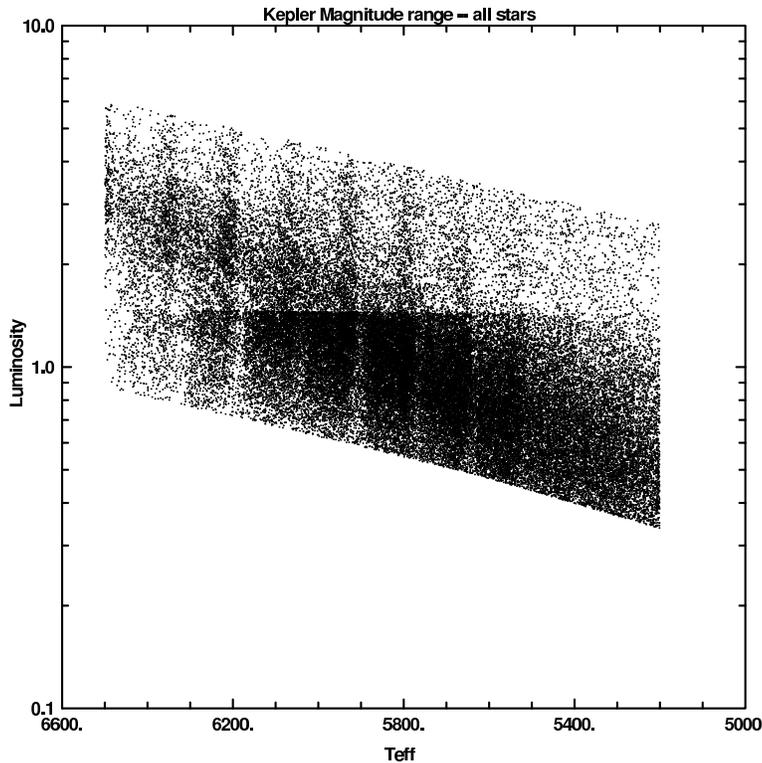}
\caption{An HR-diagram for the 69,005 stars retained for this
study after eliminating sub-giants, giants, eclipsing binaries,
KOIs, etc. as discussed in the text.
Luminosity is in solar units.
\label{fig:gm.starset}}
\end{figure}

\begin{figure}
\includegraphics[width=10cm]{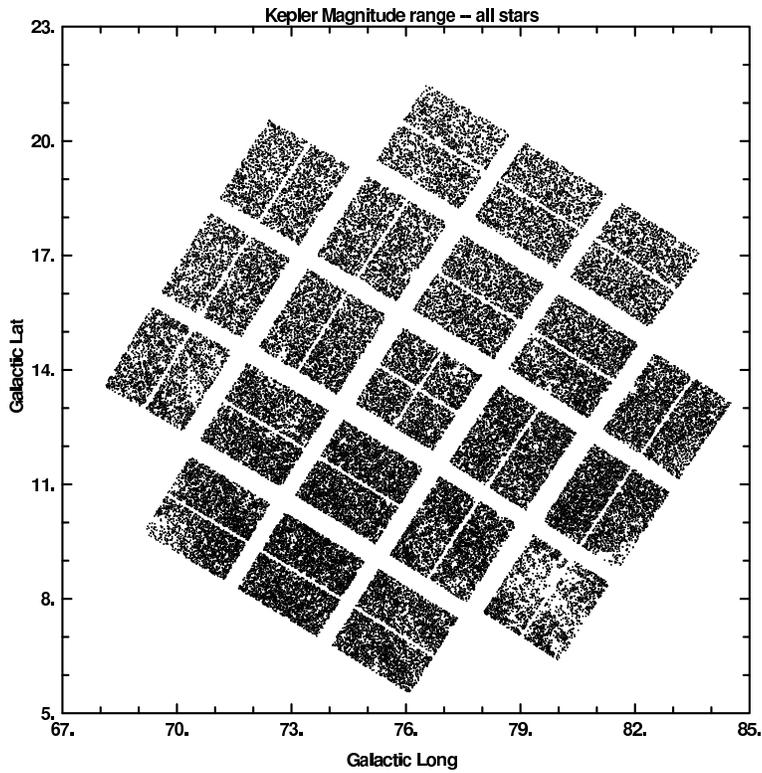}
\caption{The distribution of the 69,005 retained stars in 
galactic coordinates.
\label{fig:gm.galdist}}
\end{figure}

\begin{figure}
\includegraphics[width=10cm]{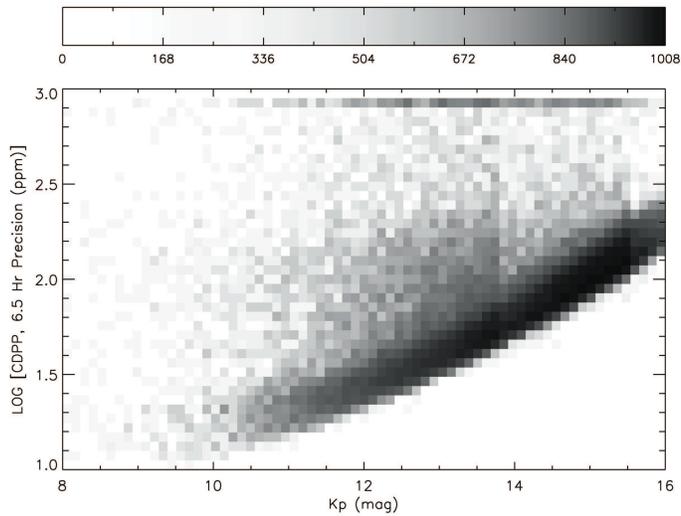}
\caption{The scatter of CDPP proxies as discussed in the text
for the 69,005 retained stars, all of which are believed to be
dwarfs of roughly solar-type based on KIC parameters.
Known binaries and KOIs were not included.  The noise values
are averaged over Quarters 2 through 6.
The band at top includes all cases at and above this level.
The bar at top defines the density of stars per 0.1 (mag) $\times$ 
0.05 (Log CDPP) bin.
\label{fig:gm.cdppo}}
\end{figure}

Fig.~\ref{fig:gm.starset} shows the distribution of retained stars
in an HR-diagram.  
Luminosity follows from the KIC mass, log g and T${\rm _{eff}}$.
The upper boundary follows from ${\rm \nu_{max}}$ = 1000 $\mu$Hz, while 
the lower boundary is set by the complementary ${\rm \nu_{max}}$ limit.
The distribution in galactic coordinates is shown in 
Fig.~\ref{fig:gm.galdist},
and the CDPP averaged over Quarters 2 through 6 is shown for this
set of 69,005 dwarf stars in Fig.~\ref{fig:gm.cdppo}.
The noise distribution in the lower panel of Fig. 1 from \citet{jen10b}
may be compared directly with Fig.~\ref{fig:gm.cdppo}.
A primary feature of the \citet{jen10b} figure (a horizontal swath
of stars variable at a few hundred ppm) is not included in our 
Fig.~\ref{fig:gm.cdppo} as a result of excluding giants and subgiants.

The CDPP, formed for each star as the mean over Quarters 2 -- 6 after
excluding the maximum value, is shown in Fig. 5 as distributions
over galactic latitude and the fractional distribution with noise level.
CDPP shows a broad distribution
with a mode of 25 ppm, with overall median 
(from set to CDPP $<$100 ppm) of 27.9 ppm.
The overall mean evaluated over a range to 3$\times$ the
median is 32.0 $\pm$ 0.3 (s.e.) ppm.

\section{STELLAR AND INSTRUMENTAL NOISE DECOMPOSITION}

The {\em Kp} = 11.5 to 12.5 magnitude range will be used for primary
analyses here. Our goal is to achieve separation of the several terms 
most important for determining the observed noise (CDPP) level
of stars.  Within the 12th magnitude band we have about 2500 
dwarf stars in our sample, most of which were observed five times
(a small subset was lost in Quarter 4 and after from the loss of
module 3, and a minor geometric distortion of the focal plane leads to stars falling
on or off silicon at the $<$1\% level
Quarter-to-Quarter), there are thus some 12,500
observables.  We wish to solve for five parameters corresponding
to any unique extra noise Quarter-to-Quarter, 84 parameters
corresponding to extra noise that may be associated with 
individual channels, and the 2,500 intrinsic stellar noise values.
Such a decomposition can be easily set up as a general least
squares problem with the normal equations having some 2,589 
unknowns and 12,500 observables.  While formally over-determined,
the problem possesses multiple degeneracies.  This forces us to use
some finesse in obtaining a decomposition.

We work primarily in variance space where we assume that
the different noise terms are uncorrelated and can therefore
simply be added.  For quoting results we will often use
noise estimated as the square root of variance.

The first step in all decompositions is to express the 
observables as variances, simply by squaring the CDPPs
available in ppm.  The next step is to remove deterministically
set terms for Poisson noise from the target and sky, and 
detector readout noise.
This combined noise term for the target is set as the 
inverse of the projected signal-to-noise ratio.
The signal term
uses the median value in e- for the PDC
(in archival time series these are the `ap\_corr\_flux' vectors) light curve
(which has had an offset ascribed to neighbors removed in 
the pipeline).  In the sum of variances
for noise the median value of the ``raw" lightcurve
(`ap\_raw\_flux' vector) is used
since the excess flux in the optimal aperture due to neighboring
stars will at the least contribute extra Poisson noise.
As an equation the combination of Poisson noise from target and sky,
plus detector readout noise is:

\begin{equation}
{\rm Noise (ppm)} = 10^6 ({\rm raw_{cts}} \, + \, {\rm readout} \,+\, {\rm sky})^{1/2}/{\rm PDC_{cts}}
\end{equation}

\noindent
where all terms are sums over appropriate numbers of exposures and pixels.
The term for detector readout variance is the product of number
of reads in 6.5 hours (13 $\times$ 270 = 3,510) multiplied
by the number of pixels in the optimal aperture for the star
in a given Quarter, and then multiplied by the square of the
readout noise as shown in Fig.~\ref{fig:gm.obs}.  The resulting
readout variance is about a factor of 10 smaller than the 
Poisson variance for a 12th magnitude star.  The contribution
from Poisson noise on the accumulated sky background is another
factor of several smaller.
The
variances passed forward for further analysis are now the observed
variances (i.e., CDPP squared) from which the Poisson and readout
noise have been subtracted.

\subsection{Quarter-to-Quarter Variations}

  Each Quarter ($\sim$90 days) of {\em Kepler} observations will be
unique.  Some Quarters, such as number 2, suffered from a 
larger number of spacecraft safing events than usual, and during this 
Quarter variable guide stars were still leading to minor 
variations of the guiding having print-through effects on
the photometry for some stars.  Finally, until mid-way through
Quarter 3 a spacecraft heater was operated in a way that
introduced minor guiding errors on a time scale influencing
CDPP; this was compounded by larger than average
focus changes occurring during Q3.  
These and other issues with the data are detailed
in the `Data Release Notes', e.g. \citet{chr10} for Quarter
2 that may be found at the MAST {\em Kepler} Archive.

  Seeking Quarter-to-Quarter mean offsets we start
by evaluating our sample of about 2,500 12th magnitude
stars to determine what their mean variance was in each Quarter.
We have also set up the full set of normal equations and 
directly solved for the Quarter-to-Quarter variance offsets
using Singular Value Decomposition codes \citep{pre92}.
The two approaches gave nearly identical results.
Table 1 records the by-Quarter variance excesses, and 
equivalent excess noise relative to the quietest Quarter (Q5),
which has been set to zero.
The resulting variations are consistent with expectations
given improved management of the spacecraft from Q2 to Q3
and beyond.

\subsection{Channel-to-channel Instrument Noise}

  We have already accounted for ordinary CCD readout noise
and always subtract this term as deterministic before searching
for more subtle contributions.  While the {\em Kepler} CCDs are
excellent for their intended purpose, a few of them suffer from
excess noise due to a number of limitations with the electronics.
The Data Release Notes discuss these \citep{chr10}.  For our
purposes we simply wish to solve for the excess noise that may
exist channel-to-channel without making reference to the detailed
reasons for such.

 Our solutions are developed on a quartet-by-quartet basis, 
e.g. the set of channels 1, 29, 53, and 81 which comprise one
of twenty-one quartets.  Over the five Quarters the same sets
of stars are seen by each of these channels in turn.   
A degeneracy exists between the global offsets between all
four such channels and a global offset for the approximately
120 stars per quartet observed each Quarter at 12th magnitude.  We can solve for 
these channel-to-channel offsets and do so using first SVD,
followed by a direct solution of the normal equations for which
the least noisy channel has been fixed to zero excess noise to
remove the degeneracy.  Doing this over all 21 quartets then
results in {\em relative} offsets for the channels within the
quartets, and we additionally track the median value (within
range of 0 to 70 ppm) of the stars for each quartet.

  Table 2 provides first a mapping of channels contained in
each of the 21 quartets, then the channel to channel offsets
found from the above prescription as well as the resulting
median for the stars in this quartet.
We will return to further discussion of this nature, but note
that channels showing unusually high excess variance correlate
at least qualitatively with channels already suspected of 
poorer than average performance.  For example, channel number 58 (in third 
channel of quartet 14) is known from ground-testing to have 
the worst overall performance in terms of image artifacts 
arising from limitations in the electronics boards, and 
the excess variance for this channel is clearly on the high
side of values in Table 2.  In a similar way the channels with
the poorest overall focus, starting with the worst are: 56, 55,
3, 53, 54, and 4 with variances of 250, 227, 247, 192, 210, and
216 ppm$^2$ respectively, thus corresponding to several of the overall
largest terms.

\subsection{Intrinsic Solar Noise}
\label{sec:intrinsol}

  We develop intrinsic noise levels for the Sun using
12 yr of measurements with the VIRGO/SPM instrument
in the green wavelength channel onboard the ESA/NASA
SOHO spacecraft (Fr\"ohlich et al. 1997).  We divided the 
1996 -- 2008 SPM lightcurve, which spans a full
solar cycle, into thirty uninterrupted bins 91.7 days
long with intrinsic cadence of 60 seconds, then binned
these into 29.4 minute sums (by averaging 30 consecutive
points with weights of 0.7 for the end points).
These sets of data were then treated in an identical 
manner using the same codes to arrive at CDPP estimates.
In the case of these SOHO data instrument
contributions to the variance are negligible and these
estimates provide a good measure of the intrinsic 
variability of the Sun on the time scales of interest
for this study.

The VIRGO/SPM green channel is at a wavelength of 500 nm,
5 nm wide.  We thus need to correct for the differing 
wavelengths of the SOHO and {\em Kepler} observations 
assuming the canonical linear scaling of amplitude with
wavelength \citep{kjeldsen95}.  The {\em Kepler} bandpass
is very broad covering 423 -- 897 nm at the 5\% points
\citep{koc10}.  Integrated over the spectral energy 
distribution for a solar-type star the mean wavelength
for {\em Kepler} observations is 634 nm \citep{van09}.
We therefore scale the
derived CDPP values from the VIRGO observations down
by 500/634 to approximate the variability for the 
{\em Kepler} bandpass.  This provides a mean and rms
of 11.0 $\pm$ 1.5 ppm, with the full range of variations
encountered over the solar cycle spanning  7.8 to 14.7
ppm -- see Fig.~\ref{fig:solarex}.  These values will be used for comparison with
the distribution of stellar noise as observed for ensembles
of stars with {\em Kepler}.

\subsection{Imposing Normalization Over Channel Quartets}

The variance normalizations appearing in Table 2 assume that
the quietest channel of each quartet has no excess variance.
Most such channels will in reality
have non-zero, and positive excess variance.  Currently 
there is a direct degeneracy within each quartet between
the value for the channel forced to zero and the overall
level of stellar variance (as given by the median over 
0 to 70 ppm) for all stars within the quartet.

The $\sim$120 stars within each quartet are usually drawn
from widely different and symmetric positions on the sky.
Therefore, even if there were intrinsic large scale (but 
low order, say a gradient over the field) 
variations of the stellar noise spatially, we should expect
the quartets to all have similar levels of stellar 
variability.  This provides a good and physically 
reasonable, albeit imperfect, means of 
renormalizing the variance levels within each quartet: 
we force the stellar medians to 
have a common value for all quartets via simple addition.  This brings the 
full distribution of channel-by-channel variances to 
a common scale.  The stellar variances are also 
brought to a common scale, although simple gradients
of intrinsic variation across the focal plane should
be preserved by this step.

The problem at hand thus becomes one of determining
what single number to adopt  
for the stellar quartet medians.  The overall minimum of the 21 
by-quartet stellar variance medians is 240 ppm$^2$, at quartet no. 2, and
may be thought of as the minimum value for this parameter,
although this choice would lead to a large fraction of 
stars with apparent negative intrinsic variance.
If instead we chose this value to be the overall mean of 378 ppm$^2$
for the stellar medians over all 21 quartets,
then quartets 2, 5, and 13 with low values of the 
stellar median would develop many channels with
negative variance for the instrument noise contribution.

There does not seem to be a well-defined quantitative 
approach to selecting this normalization.  We will
rather rely on showing the sensitivity to this choice
and bringing in congruence with solar variability 
to support our ultimate selection.

Fig.~\ref{fig:gm.3hist} shows the fractional distribution of
noise values with three different choices of the normalization
for the stellar median, from a minimal value of 300 ppm$^2$
at top to 378 ppm$^2$ at the bottom.
Such changes in the overall variance zero point between possible
instrument terms and noise intrinsic to the stars makes 
a noticeable difference for the really quiet stars, but has much 
less influence on stars with higher intrinsic variability.

The distribution of stellar noise seems to be bi-modal 
in character, with one peak in the neighborhood of typical
solar variations, and a second peak at some 5 ppm 
higher values.  In addition an extended tail to higher 
intrinsic noise exists for the more active stars.

We adopt the 340 ppm$^2$ variance level for the median
intrinsic stellar noise over 0 -- 70 ppm.  Having done 
this we may now present the results of Table 2 in a different
way.  Values in the column ``Stellar" would now all be 340
by design.  Fifteen of the twenty-one quartets have 
stellar median values above 340, for these the by-channel
variances will be shifted to more positive values as
the difference between these entries and 340.  There are
six quartets for which the stellar median is smaller than
this target of 340 ppm$^2$ which has the consequence 
that at least the channels starting with zero variance 
will shift to non-physical, negative variance values.

To assess how concerned we should be by finding a few
channels coming in at negative additional variance it 
is useful to determine the noise levels that apply to 
the by-channel variance numbers in Table 2.  We have
done this by opening the magnitude range up slightly
from 11.5 -- 12.5 to 11.2 -- 12.8 {\em Kp}, thus doubling
the number of available stars.  Then we have repeated
the analyses leading to Table 2 independently for 
stars with even or then odd Kepler Input
Catalog (KIC) numbers.  As expected this led to two
sets each about as large as the number of stars in the
Table 2 sample.  The gross features of derived values
in these even and odd sets are very similar; channels
standing out as high in Table 2 are high in each of the
even and odd test sets, as they are for the distribution
of stellar medians by quartet.

The overall rms across the variances between the even
and odd sets, after shifting all for consistency with
a common stellar median of 340 ppm$^2$, is 60 ppm$^2$.
The estimated standard deviation on the direct values
in Table 2 would be a factor of two smaller than this,
or 30 ppm$^2$ as a general confidence level estimate.
Following correction to a common stellar median of 
340 ppm$^2$ in Table 2 entries a total of 12 channels
have formally negative variances, and 5 of these are
more negative than the nominal 1-$\sigma$ level.

The number of channels coming in with negative 
variance does not seem inconsistent with expected 
fluctuations given our measurement precision 
channel-to-channel.  Similarly, in Fig.~\ref{fig:gm.3hist}
at the nominal 340 ppm$^2$ median value
some 4\% of the stars overall have entries in the first
bin which corresponds to those with a negative variance.
Stars with negative variance may arise
from any channels that have significant gradients of 
intrinsic noise.  Then a quiet star, on the less noisy 
part of the channel could be driven to negative variance
when the correction for the channel as a whole is applied.
Also, in some cases PDC will over-fit and remove actual
noise, but this is not thought widespread.

The existence of a small fraction of channels ascribed
a negative intrinsic variance, and a few stars determined
to have intrinsic variances less than zero, both of which are 
non-physical, is a signal that our
determinations of these quantities introduces 
some broadening of intrinsic distributions.

\subsection{Dependence of Stellar Noise on Galactic Latitude}

The overall results for intrinsic stellar noise for the
12th magnitude sample may be seen in Fig.~\ref{fig:gm.cdppdist}.
The noise has a moderate dependence on galactic latitude,
and apparently bi-modal peaks near 11 and 17 ppm, with a 
tail to larger values.
From plots (not shown) of stars within 7--11 ppm bands,
and 17-21 ppm bands from the histogram of Fig.~\ref{fig:gm.cdppdist}
the stars appear to be uniformly distributed over the full 
focal plane.  There is a mild dependence on galactic latitude
in that noisier stars are statistically more likely at 
low galactic latitude:  for intrinsic noise of 30--70 ppm 60\%
are at $b$ $<$ 13 degrees, while at 17--21 ppm and 7--11 ppm
the fraction drops to 49\% and 45\% respectively.
The medians (up to 100 ppm) and means (using stars up 
to 3$\times$ median) over all latitudes are 19.6 
and 20.3 $\pm$ 0.3 ppm respectively.
A more realistic estimate of the errors for the median
and means would be to adopt the spread resulting from
the range plotted in Fig.~\ref{fig:gm.3hist} which yields
$\pm$ 1.0 and $\pm$ 1.3 ppm respectively.

Representative stars from quiet, moderate and high noise
levels are shown in Fig.~\ref{fig:gm.3exam}.  For the 
quietest stars the {\em Kepler} photometry does not definitively
show any variability.  By the time stars
have intrinsic noise levels over 30 ppm it is common to be
able to see evidence of what is likely rotational modulation as seen in
the bottom panel of Fig.~\ref{fig:gm.3exam}.

\subsection{Consideration of Crowding and Superposition on Noise}

We have considered contributions to variations in detected
flux arising from Poisson statistics on the source and sky,
detector readout noise, separate additive (in variance) terms
that vary by Quarter, and by detector channel number.
After accounting for the above, residual variations have so far been assumed to represent
changes inherent to the individual stars.

There are two additional terms that need to be considered, in
at least a statistical sense, to discern if the overall
decomposition of noise terms is reasonable.  Given the 
large pixel scale of {\em Kepler} it is common for significant
amounts of light within the photometric apertures to arise
from nearby neighbors -- the amount of this that we know 
about is referred to as the contamination fraction.
By design we have limited stars in our {\em Kp} = 11.5 to 12.5
sample to have less than 20\% of the light provided by
neighboring stars (fewer than 1\% of 12th magnitude stars
observed by {\em Kepler} are above this 20\% cut).
In evaluating the Poisson statistics
contribution we have not included this maximum of 20\%, average of 2\% in the signal,
but have allowed for this to be included in the statistical 
noise.  We have not, however, allowed for potential 
variability of the contaminating stars introducing modulation
of the target star flux.  In most cases the DC level of
contaminating fluxes are known from relative magnitudes,
separations, and assuming a PSF, but the contaminating 
star will not typically have been observed by {\em Kepler}.
This may be thought of as a ``known unknown"; we know the DC component,
but not the AC contribution in the combined time series.
Of the ``unknown unknown" variety there will be a chance
that target stars have background stars within the 
photometric aperture.  We will have already accounted
for any Poisson statistics provided by such unknown 
blended stars, but will not have allowed for possible
variations in the signal flux provided by these stars.

In both of these cases we can statistically estimate
how much the CDPP and inferred intrinsic stellar noise
is likely to be influenced by
these added contaminants.  We address the issue of 
blended background stars first.
We adopt several assumptions that will if anything 
over-estimate the influence of blended background stars:
\begin{enumerate}
\item The aperture size, or area on the sky within which
to consider statistical blends is taken to be 
the actual optimal aperture size
with a half-pixel radius buffer.
At 12th magnitude this padded aperture averages 19.6 pixels,
but is allowed to vary star-by-star and Quarter-by-Quarter
to match actual values used in the pipeline.
\item We adopt the set of {\bf all} {\em Kp} = 11.5 to 12.5 stars
that were observed in each of Quarters 2 -- 6 as a set 
from which to randomly pull time series to be added 
with appropriate dilution depending upon relative magnitudes.
Since red giants are inherently much noisier than dwarfs,
and many of these have been dropped \citep{bat10} from the
planetary target list, we replicated the observed red giants
to bring their total fraction up to 50\%.  This is slightly
greater than the 47\% red giant fraction implied by the 
KIC at 12th magnitude, and likely an 
even greater over-representation for much fainter background stars.

\item Star counts as a function of galactic latitude over the {\em Kepler}
field were adopted from the Besancon galactic model
\citep{rob03}, of which over {\em Kp} = 14--16 were some 20\%
larger than an estimate based on the KIC directly at field center.
Over the 19.6 pixel aperture the number of stars per
magnitude bin was adopted as tabulated in Table 3 for three galactic latitudes.
\end{enumerate}

To simulate the impact of unknown background stars, that if sufficiently
variable would increase the apparent variability of the usually much
brighter target star, we start with the 69,005 set of time series.
For each star the probability of a background star within one magnitude
bins over {\em Kp} = 13.5 to 22.5 overlapping with the target aperture is assessed.
For example at $b$ = 13 degrees there would be a 14\% chance of a
contributing star at 17.5, and 52\% at 22.5, etc.
When the random selection probability indicates a contaminating 
background star should be added, a random draw from the set in number
2 above is made, diluting the added signal as the ratio of brightness
of the background bin to the source star.  When making this addition
the similarly scaled median of the source star is subtracted so that
we effectively add in just the variability, and not a zero point offset
in total counts. 
Only about 10\% of {\em Kp} = 12 stars based on Table 3 would have a star
within a $\delta$-magnitude of about 4 contributing, but summed
to a $\delta$-magnitude of 10 the number of stars added in averaged
about two at field center.  When the number of expected stars exceeds
unity per magnitude bin a random draw will always be added in with appropriate 
dilution based on magnitudes, but also scaled back up by the 
effective number of such stars, e.g. $\times$ 2.22 at low galactic
latitude and {\em Kp} = 22.5.

The resulting time series are then run through the same software
used to extract the proxy CDPP values, and that is then used to decompose
the noise into the several terms discussed earlier.
The primary result of allowing for additional blended background
stars is shown in Fig.~\ref{fig:gm.simback}
where the relative effect on stars having 5 -- 20 ppm from Fig.~\ref{fig:gm.cdppdist}
is shown.  The majority of originally quiet stars are perturbed by 
less than 1\%, with about 10\% increased in inferred noise level
by 1\%, etc.  The fraction of stars originally within the 5 -- 20 ppm
domain that moved into the 40 -- 70 ppm range total only 1.5\%, and 
the number fraction from this quiet set moving to greater than 
70 ppm is 1.3\%.  While not entirely negligible, the impact of 
allowing for variable background stars is minor.  The number of 
stars promoted to the 40 -- 70, and greater than 70 ppm domains
is about 10\% of the number of stars observed to be this noisy.
The overall increase of median and mean noise evaluated
as in \S 3.5 are only 0.2 and 0.1 ppm.

The second consideration is to assess by how much the known levels
of contamination are likely to contribute for formal estimates 
of intrinsic stellar noise.
For each star we take the maximum contamination (fraction of light
within optimal aperture contributed by known neighbors) over Quarters
2 - 6 for the {\em Kp} = 11.5 to 12.5 sample, the mean of this is 0.021.
Using a similar approach as for the background stars, we take the
contamination fraction as the dilution factor (adjusted as well
for the relative brightness of target and randomly drawn perturber),
and always add in one randomly drawn additive star.
The results are shown in Fig.~\ref{fig:gm.simback}, again as the
relative impact on stars that were within 5 -- 20 ppm intrinsic
noise before this.  A larger fraction of stars are moved to 
slightly higher noise levels than was the case accounting for
background stars.  However, as measured by the number promoted
from the 5 -- 20 ppm bin into, or beyond the 40 -- 70 ppm range
the fractions remain very modest at 0.9\% and 1.2\% respectively.
The fractions of stars at high noise levels seems perturbed 
by about 10\% after considering the effects of known contaminating
stars and allowing for the fact that some of these will be 
significantly variable.
The overall increase of medians and mean relative to the 
base from \S 3.5 are again negligible.

The estimates in this section cannot be used to correct 
inferred stellar noise levels on a star-by-star basis, but 
do indicate that the impact of either known
contaminating stars, or fainter, blended background variables
is a statistically minor factor overall.  Summaries here and in 
Fig.~\ref{fig:gm.simback} have considered only statistically
average responses over all channels, and at all galactic
latitudes.

The equivalent of Fig.~\ref{fig:gm.simback} drawn
for only low galactic latitudes would show somewhat larger
responses.
We quantify this by evaluating the slope of the mean values
over 1 degree steps for $b$ = 8 to 18 for the equivalent of
Fig.~\ref{fig:gm.cdppdist} for these simulations.
The direct observations have a slope of -0.58 ppm/degree,
while the background and contamination simulations come in
at -0.81 and -0.67 respectively.
Combined, the excess slope from these simulations is -0.32
ppm/degree, suggesting that about half the apparent galactic
latitude dependence may result from contamination effects.
Residual evidence for intrinsic stellar variability 
dependence on galactic latitude is weak.
We will return to these higher order
considerations in \S 3.8.

\subsection{Overall Noise Decomposition Results}

Having now derived noise terms covering contributions 
that change in time unique to each Quarter, that arise
from the detector channels, assigned Poisson and readout
noise, and intrinsic stellar noise, we may now examine
this from a global perspective.  In particular, what is 
the ordering of terms and how do these compare with
expectations?  If we now put the decomposed terms back
together is the CDPP effectively reproduced?

Table 4 provides the four independent variance terms
into which we have decomposed the observed variance
(CDPP$^2$).  The largest of these is the intrinsic 
stellar variability for which our value is 19.5 ppm.
The budget for stellar variability had been set 
to 10 ppm \citep{jen02}; since this term is both the
largest contributor overall, and twice the budgeted
value, this term strongly deviates from expectations 
in a significant way.  {\bf  The largest contributor
to higher than expected CDPP is an intrinsic stellar
variability significantly above expectations.}
The second largest term at 16.8 ppm includes 
contributions from Poisson noise on the target star
itself, and much smaller contributions from Poisson noise 
of the sky, and detector readout noise.  
The mean {\em Kp} of this 11.5 -- 12.5 sample of stars is
12.10, and the irreducible terms for Poisson noise
and readout are entirely consistent with expectations.
The third largest term at 10.8 ppm is a global 
average (as variances) over the 84 detector channels.  Since the 
detector noise contribution is well below the 
level of Poisson noise this has little influence on
the bottom-line CDPP.  Expectations were that a large
number of imperfections in the electronics would 
combine to provide several ppm of equivalent noise,
so what we find is broadly consistent with expectations,
albeit marginally higher.  Finally, the Quarter-to-Quarter
component of noise comes in at 7.8 ppm.  This had not 
been explicitly budgeted for in CDPP estimates, and
this term is dominated by Q2 and Q3 which both had
excess noise from safing events, telescope repointing,
and minor heater cycle/guiding issues that have since
been brought under control.  Averaged over Qs 4--6
this time-dependent term would be under 5 ppm.

The final line of Table 4 shows the sum of the 
four variance terms, and derived overall CDPP for 
the {\em Kp} = 11.5 to 12.5 sample at 29.0 ppm.  This is 
quite consistent, as it must be unless an inconsistency
had arisen with the decompositions with a consistently
derived CDPP value for the same stars.
This value is 50\% larger than the target of 20 ppm.
Had intrinsic stellar variations held to 10 ppm for
a large fraction of the stars, instead of adopting a
broad distribution extending beyond 20 ppm, then the
noise roll up would be at 23.6 ppm.  (Part of this 
would be accounted for by our sample mean magnitude
being fainter than 12.0.  A caution applies, though, that scalings
in either magnitude, or time don't have precisely the simple form
expected due to a varying number of pixels as a function of 
magnitude and the stellar noise being red respectively.)
Without the higher intrinsic
stellar variability we are some 10\% above
the requirement of 20 ppm CDPP at {\em Kp} = 12, with
the higher intrinsic noise from stars factored in 
we are some 50\% above our noise requirement.

\subsection{Detector Noise by Channel and Focus Dependence}

It is now instructive to document the inferred excess
noise by channel and the dependence of this excess on 
the instrument focus, which varies significantly over the
focal plane.
Fig.~\ref{fig:gm.chnoise} is in the same style as Fig.~\ref{fig:gm.obs}
and shows the excess noise level as derived following the
steps in \S 3.2 and 3.4 above.  
As argued in Table 4 the overall contribution of this 
excess channel noise is rather modest for the CDPP error budget.
Nonetheless it is interesting to note that patterns in this
noise exist, and seek at least a correlative explanation.

Fig.~\ref{fig:gm.varfoc} shows the intrinsic 
excess variance by channel plotted versus the fractional central pixel energy content
for a star centered on a pixel for that channel.
The straight line is a simple least squares fit of 
variance with the focus proxy, assuming no error in the 
latter.  The linear correlation coefficient of excess
variance with focus is 0.63, with a sign that clearly
indicates the better photometry follows from channels
with best focus.  This is true even when the resulting
PSFs are severely undersampled.  As argued in \S 2 the
exquisitely good guiding provided by {\em Kepler} supports
excellent photometry even for sharp PSFs; indeed it does
best in this circumstance.
This convincingly argues that there is no significant
effect from intra-pixel sensitivity variations coupled with jitter.

In \S 3.6 we demonstrated that the large {\em Kepler} 
apertures, which will lead to blended background (or 
foreground) stars, do not lead to significantly changed 
overall noise statistics when these blends are simulated.
For a channel with soft focus (low focus values in
Fig.~\ref{fig:gm.varfoc}), larger apertures are used
to capture the target flux.  This will lead to both 
larger values for the known contamination from nearby
neighbors on the sky, and higher probabilities that
unknown blended objects are within the aperture.
Using the simulations performed in \S 3.6 we have 
independently produced versions (not shown) of Fig.~\ref{fig:gm.varfoc}
for both the known and unknown background contaminating
cases.  In both instances the gross features 
of Fig.~\ref{fig:gm.varfoc} are well preserved, although
in both cases the slope of channel variance with focus
increased by about 10\%.  This implies that some 20\%
of the implied channel noise dependence on focus results
from the inclusion of more variable stars in poor focus
channel apertures.  (There is some double counting
between the known and unknown simulations, but since the
perturbation for both combined remains small we have not
attempted to quantify this.)

Further support for the reasonableness of the noise 
decomposition comes through comparison of entries in Fig.~\ref{fig:gm.varfoc}
for channels showing moir\'{e} patterns in the background
based on ground-based, and Q0-Q1 data analyses \citep{cal10,kol10}.
Of the ten channels having worst moir\'{e} features, eight of these
fall above the linear fit with a mean offset of 72 and standard
error of 21 ppm$^2$ -- a significance of over 3 $\sigma$.
Seven of nine channels flagged independently as moderate in their 
moir\'{e} features are above the linear fit, but the mean offset
drops to 28 $\pm$ 31 ppm$^2$.  
Four of the five channels with largest excess noise in Fig.~\ref{fig:gm.varfoc},
either in absolute terms, or relative to the linear fit correspond
to those flagged for moir\'{e} influence.
The moir\'{e} pattern noise is thought to arise from temperature-sensitive
operational amplifiers used extensively in the video chain
electronics, which may show subtle layout-dependent instability
when used at the low gains adopted for {\em Kepler}.
The oscillation's frequency range,
rate of change, and pattern among the
channels matched closely those
characteristics in the dark images, strongly suggesting that the artifact
follows from sampling the $>$1 GHz amplifier oscillation at the 
3 MHz serial clocking rate.  Since the effect is temperature
sensitive the pattern will drift in time, and the relative importance
of the imposed moir\'{e} pattern noise will be different Quarter-to-Quarter
for the channels sensitive to this.

\begin{figure}
\includegraphics[width=10cm]{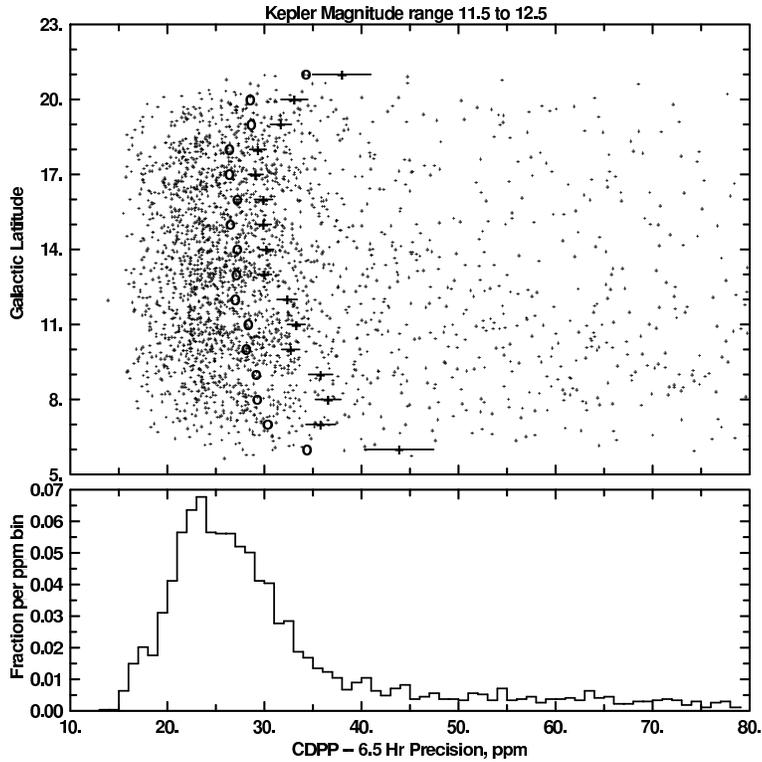}
\caption{The upper panel shows the CDPP in
ppm for the {\em Kp} = 11.5 to 12.5 sample as a function of galactic
latitude.  Medians evaluated over up to 100 ppm are shown as `o',
while means from up to 3$\times$ the median at each degree of
galactic latitude are shown as `+' symbols.  Standard errors
for the means are shown.  The lower panel shows a histogram 
of fraction of stars per ppm bin.
\label{fig:gm.ocdppdist}}
\end{figure}

\begin{figure}
\includegraphics[width=10cm]{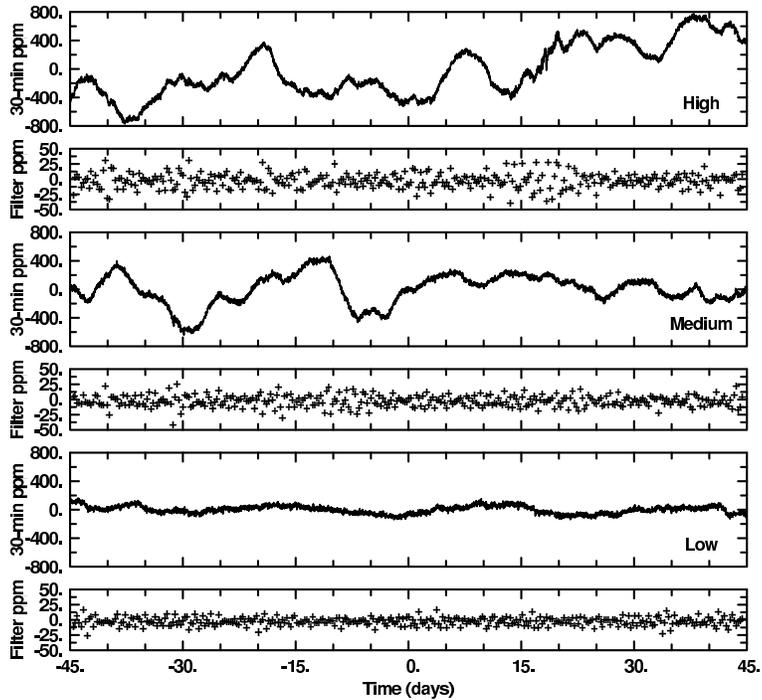}
\caption{The solar data from VIRGO/SPM
\citep{froh97} are shown for three
90 day blocks as variations in ppm around
a median. The upper panel in each shows
the direct data after binning to {\em Kepler's}
29.4 minute cadence and scaling by $\times$0.8 as
a color correction as discussed in the
text. The companion panels show the data
after applying the Savitsky-Golay filter
followed by 6.5 hour binning used to form
CDPP. The upper panel pair shows the
noisiest 90 day period centered on 2002.39
(CDPP = 14.7 ppm), the middle panel pair
is centered on 2005.52 (CDPP = 10.9 ppm),
while the low noise case (CDPP = 7.8 ppm)
is from 2007.77.
\label{fig:solarex}}
\end{figure}

\begin{figure}
\includegraphics[width=10cm]{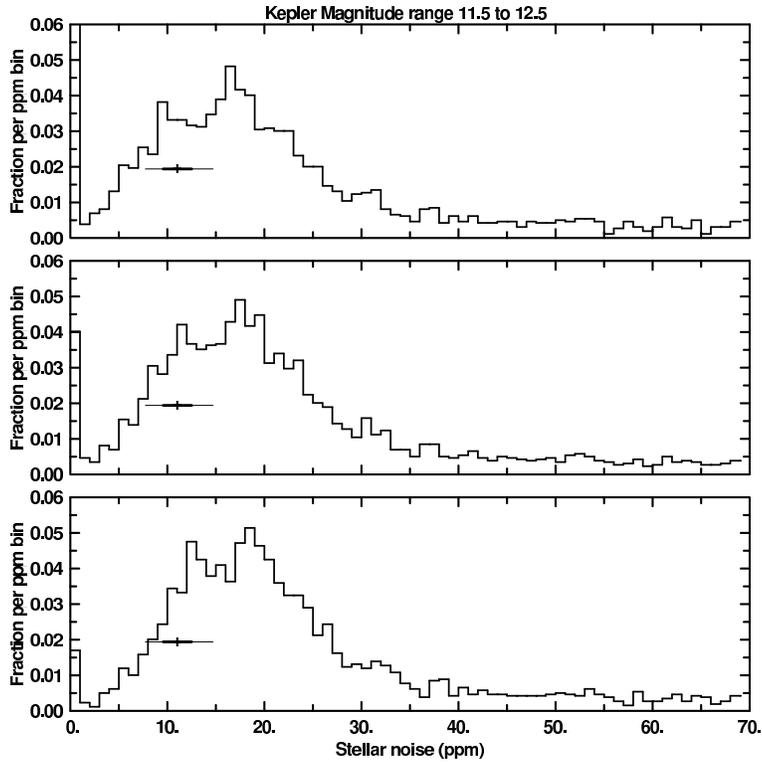}
\caption{Each panel shows the fractional distribution of 
{\em Kp} = 11.5 to 12.5 stars within 1 ppm bins for instrinsic
stellar noise.  The mean and rms distribution for solar
noise levels over Quarter-long intervals spanning a 
solar Cycle are shown by the `+' and heavy horizontal line,
with the full extent of solar noise per Quarter the thin
line.  From top the quartet stellar variance medians
are set to the minimal value (see text) of 300 ppm$^2$,
the 340 ppm$^2$ value we adopt as most representative,
and the bracketing 378 ppm$^2$ on the high side at the bottom.
\label{fig:gm.3hist}}
\end{figure}

\begin{figure}
\includegraphics[width=10cm]{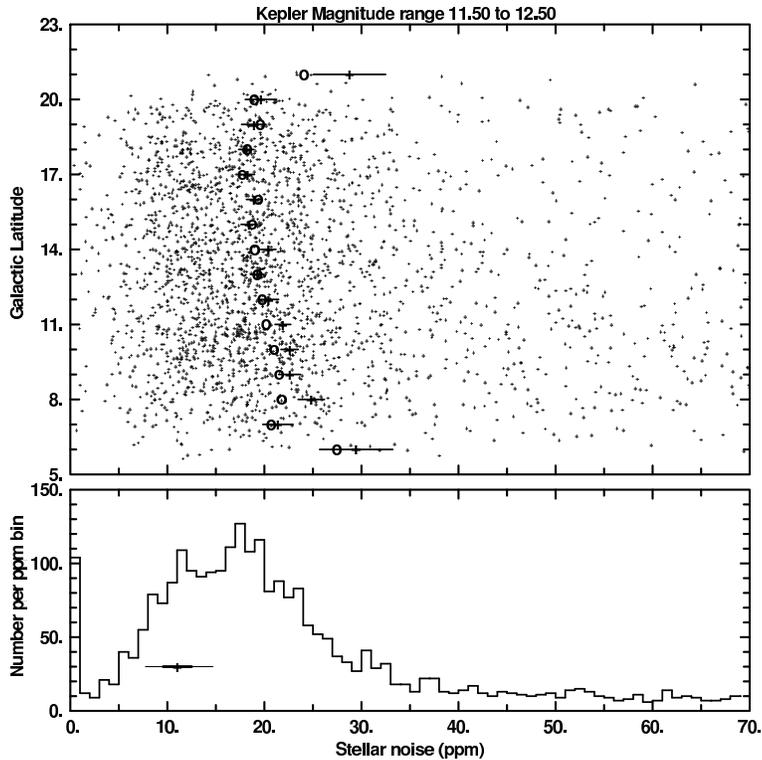}
\caption{The upper panel shows the intrinsic stellar noise in
ppm for the {\em Kp} = 11.5 to 12.5 sample as a function of galactic
latitude.  Medians evaluated up to 100 ppm are shown as `o',
while means from up to 3$\times$ the median at each degree of
galactic latitude are shown as `+' symbols.  Standard errors
for the means are shown.  The lower panel shows a histogram 
of number of stars per ppm bin.  The mean and rms distribution
for solar noise levels over Quarter-long intervals spanning a
solar Cycle are shown by the `+' and heavy horizontal line, 
with the full extent of solar noise per Quarter the thin line.
\label{fig:gm.cdppdist}}
\end{figure}

\begin{figure}
\includegraphics[width=10cm]{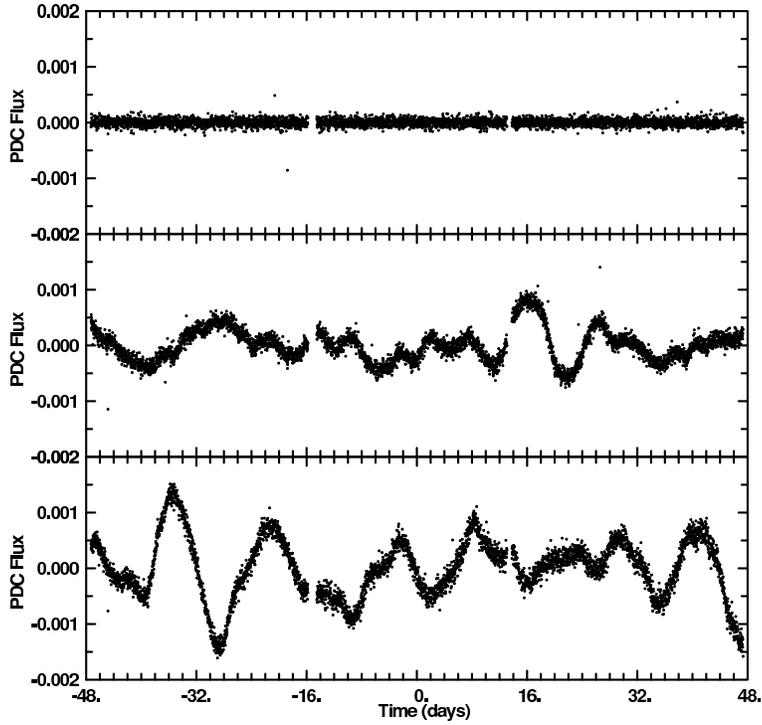}
\caption{Representative time series for stars
from low (5.7 ppm), medium (20.4 ppm), and high (35.0)
intrinsic noise levels are shown.  From top to bottom these stars
are KIC 4662814, KIC 4663537, and KIC 4283320.  All of the 
time series are from Q5 and are the direct pipeline 
provided PDC fluxes after subtracting out, and normalizing
by an overall median.
\label{fig:gm.3exam}}
\end{figure}

\begin{figure}
\includegraphics[width=10cm]{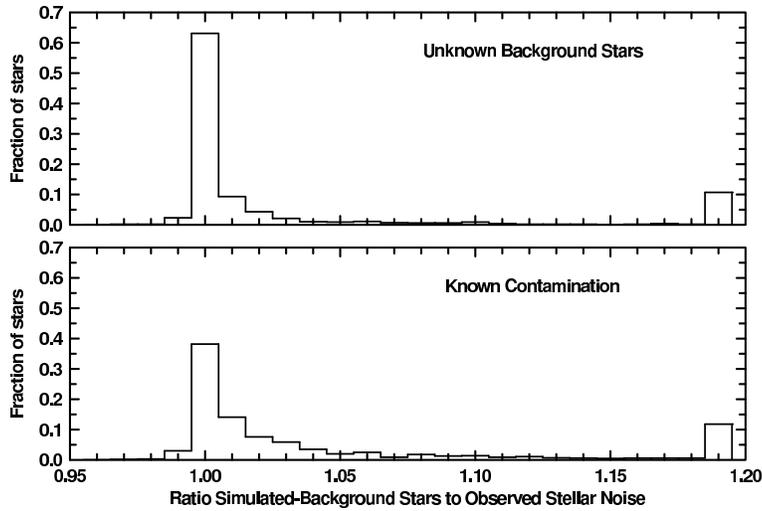}
\caption{The upper panel shows the ratio of inferred noise
levels for stars originally within 5 -- 20 ppm noise levels
to the results from adding simulated (unknown) background
variable stars.  The final bin includes all values out of range.
The lower panel shows the same for the simulations taking 
into account that known blended stars may be variable.
\label{fig:gm.simback}}
\end{figure}

\begin{figure}
\includegraphics[width=10cm]{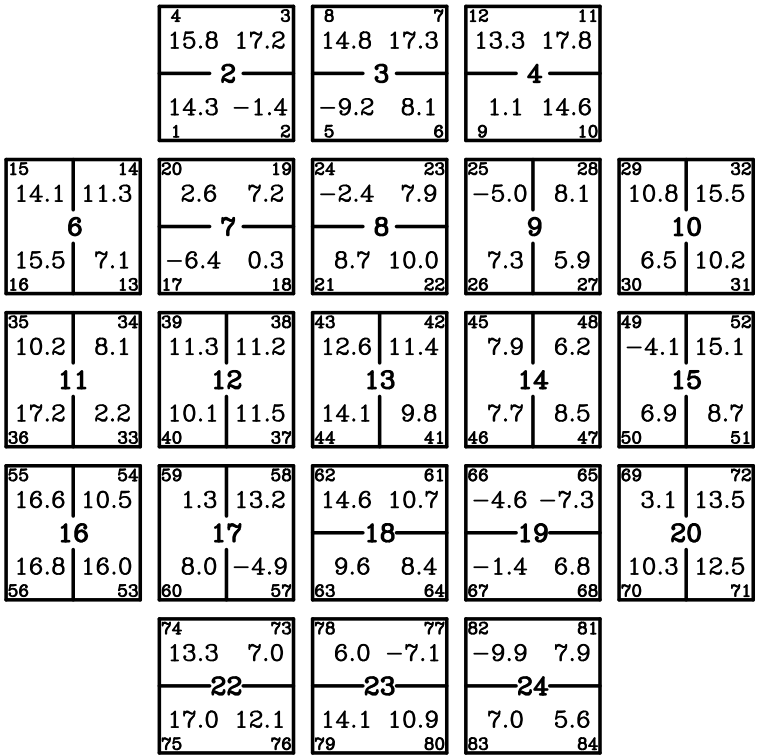}
\caption{This shows the distribution of noise intrinsic to
each detector channel derived as in \S 3.2 and 3.4.  
Solutions are in variance space, translated to equivalent 
noise by taking a square root.  Where the variance, due to 
inherent scatter in derivation, is negative, the noise is
evaluated as minus the square root of the negative of this
variance.  
\label{fig:gm.chnoise}}
\end{figure}

\begin{figure}
\includegraphics[width=10cm]{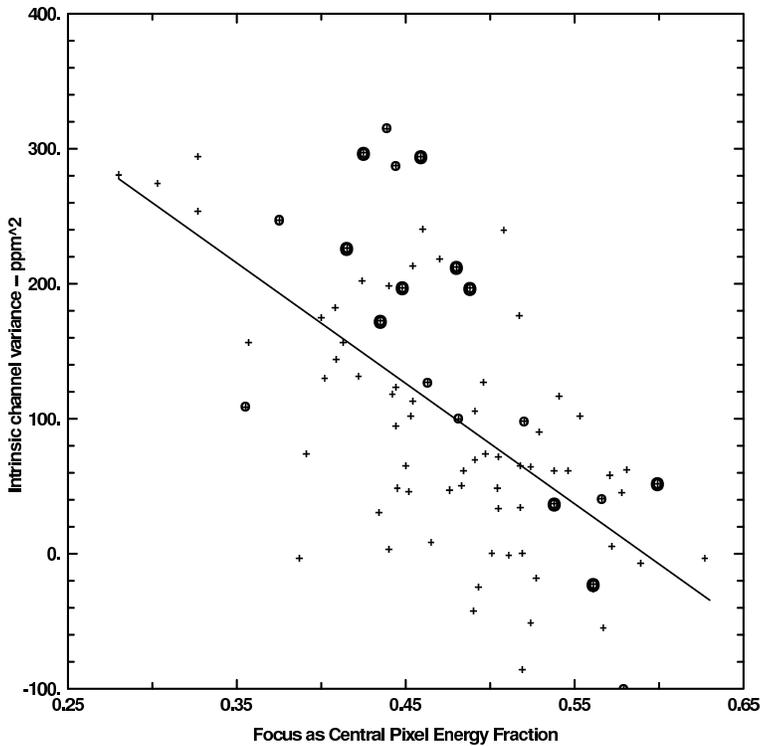}
\caption{The by-channel intrinsic variance levels are 
plotted against by-channel focus as represented by the 
fraction of total energy in the central pixel for a 
star centered on a pixel.  As expected now for {\em Kepler},
but contrary to common knowledge, the best photometry is 
returned for the sharpest PSF channels, even when the
PSF is severely undersampled.
Channels over-plotted with a small circle represent nine
cases independently identified to have moderate moir\'{e}
pattern noise, and the ten cases with strong moir\'{e}
noise have doubled circles added.
\label{fig:gm.varfoc}}
\end{figure}

Having shown that some Quarters (Q2 in particular), and channels are
noisier than others we tried creating a new version of Fig.~\ref{fig:gm.ocdppdist} in
which only the best Quarters, Q4 - Q6, and the lowest noise (29 of 84)
channels were allowed to contribute. The resulting figure was very
similar to Fig.~\ref{fig:gm.ocdppdist}.  The values going
into Fig.~\ref{fig:gm.ocdppdist} were based on means
over all Quarters, but without using the maximum (usually Q2). The
excess variance contributed by the worst channels on the instrument
remain below the level of intrinsic stellar noise, and only comparable
to Poisson contributions as detailed in Table 4. Thus the CDPP shown
in Fig.~\ref{fig:gm.ocdppdist} is considered robust.

\subsection{Stellar Correlations with Effective Temperature}

At the suggestion of the referee we have explored the hypothesis that
stars of spectral type A to early F with insignificant surface
convection zones should frequently be low noise. This is pursued in
the spirit of a sanity check on the primary conclusions of this paper,
namely that solar type stars are typically noisy (where by {\em Kepler}
standards this means ~20 ppm on timescales of 6.5 hours). To simplify
interpretations we tabulate the fraction of stars in spectral type
bins spanning A0, A5, . . . M0 that have noise levels (CDPP) less than
10 ppm for stars with Kp = 7-11 where Poisson noise contributions
allow such low values if the stars are inherently quiet, and if the
instrument is not adding significant noise. All known eclipsing
binaries and planet transit hosts have been excluded to avoid cases
with spuriously high CDPP from non-stellar events. We also show in
Table 5 what fractions of stars within these bins are quite variable
at greater than 50 ppm. The requirement is that these limits be met in
at least two of Quarters 3-6, and generally there is excellent
consistency across quarters for either quiet or noisy stars.

The primary hypothesis is well satisfied: a significant fraction of
early type stars do have noise levels that are low compared to those
that occur for solar type stars. In particular this may be taken as
support that {\em Kepler} itself does not impose a floor to the noise
levels, at least not at a level generally relevant to CDPP for 12th
magnitude stars. When the Poisson term is forced low by selecting
bright stars, and stars are inherently quiet, {\em Kepler} returns low noise
levels. There are clearly trends in the distribution of both quiet and
noisy stars with spectral type that call out for further astrophysical
interpretation, this is beyond the scope of this paper. Here, we
merely note that A to early F stars do show frequent very low noise
levels, in stark contrast to the absence of such quiet stars in the
solar type regime (with a possible small subset of quiet stars near
K0). Balancing this, though, the fraction of stars that are really
noisy as measured by our cut at $>$ 50
ppm reaches a distinct minimum for
solar type stars. The small sample
size and potential for contamination by
miss-classified giants as dwarfs
suggests caution should be exercised in
interpreting the K5 and M0 bins of
Table 5.

\begin{figure}
\includegraphics[width=10cm]{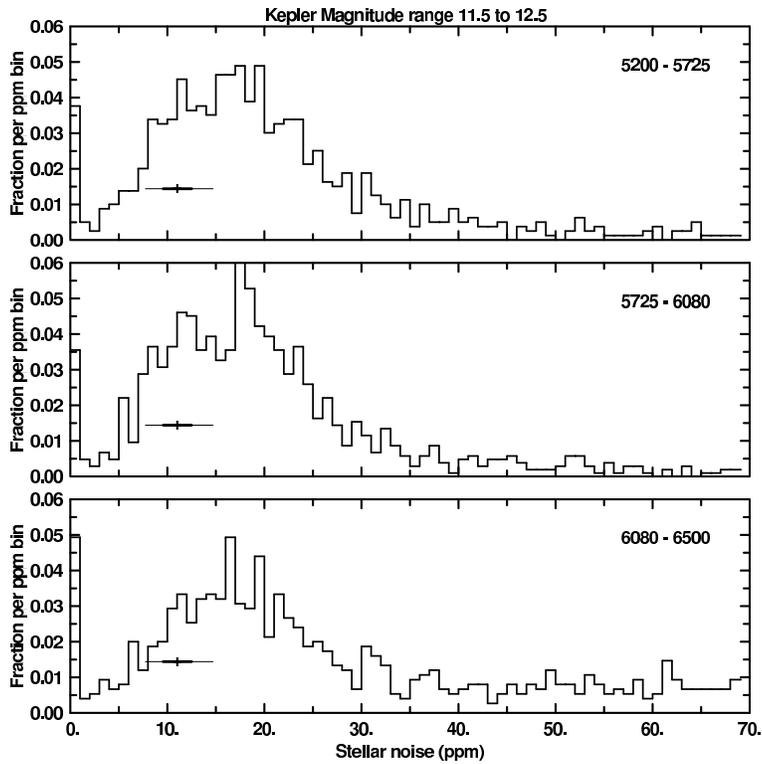}
\caption{The histogram shown in
Fig.~\ref{fig:gm.cdppdist} is here split into equal thirds of
cool (from top) to warm stars. Over the primary range of solar type
stars relative differences are subtle with the most noticeable
feature a more pronounced tail to higher values for warm stars.
\label{fig:specnrange}}
\end{figure}

The distribution of intrinsic stellar
noise within our primary range of 5200
-- 6500 K is shown in Fig.~\ref{fig:specnrange}.  The
primary difference noticeable over this
temperature range is a higher tail of
moderately variable stars for the
earlier spectral type.

\section{SIMULATION OF STELLAR NOISE}

\label{sec:sim}

We estimated the properties of the stellar population observed by
\textit{Kepler} using the code TRILEGAL \citep{girardi00}, which is
designed to simulate photometric surveys in the Galaxy.  A synthetic
population of solar-type stars was extracted for the \textit{Kepler}
field of view by applying the same selection cuts as were applied to
the real stars. From the properties of each solar-type star we then
estimated the stellar contribution to its CDPP, using the procedures
outlined in Section~\ref{sec:est} below. Before discussing these
procedures, we begin with a brief description of TRILEGAL.

\subsection{Galactic population synthesis models}
\label{sec:popsyn}

In TRILEGAL, several model parameters (such as the star-formation
history and the morphology of different galactic components) were
calibrated to fit Hipparcos data for the immediate solar neighborhood
\citep{perry97}, as well as star counts from a wide area (with 2MASS;
\citet{cutri03}), and a few deep photometric surveys, i.e., CDFS
\citep{arn01}, and DMS \citep{osmer98}.  We adopted the standard
parameters describing the components of the Galaxy and simulated the
stellar population in the sky area observed in each of the 21
five-square-degree \textit{Kepler} sub-fields of view, considering for
each of them an average interstellar extinction at infinity
\citep{schlegel98}.  The extinction is assumed to be caused by an
exponential dust layer with a scale height above and below the
galactic plane, equal to 110\,parsec.  The photometry in TRILEGAL was
simulated with the known wavelength response function of
\textit{Kepler}, and the synthetic population was magnitude-selected,
using the same range as the observed sample.

\subsection{Estimation of stellar noise}
\label{sec:est}

The stellar noise was assumed to have two components: one due to
activity, and another due to granulation. It was assumed that any
significant contribution from p-mode oscillations would have been
removed by the strong band-pass filtering implicit in construction of
the CDPP (see below).
Empirical scaling relations were used to predict parameters describing
the activity and granulation, which in turn were used to estimate the
underlying power spectral density (as a function of cyclic frequency
$\nu$) due to each contribution. We assumed that the contributions
could be described as exponentially decaying functions in time
parameterized by an RMS amplitude, $\sigma$, and a timescale,
$\tau$. The underlying power spectral density (PSD) is a
zero-frequency-centered Lorentzian of the form \citep{harvey85}
 \begin{equation}
 {\rm PSD}(\nu) = \frac{2\sigma^2\tau}{1+(2\pi\nu\tau)^2},
 \label{eq:psd}
 \end{equation}
with the variance in the time domain being equal to $\sigma^2/2$. We
discuss the validity of using this function in
Sections~\ref{sec:estact} and~\ref{sec:estgran} below.

\begin{figure}

 \centering
 \includegraphics[width=0.75\textwidth,clip]{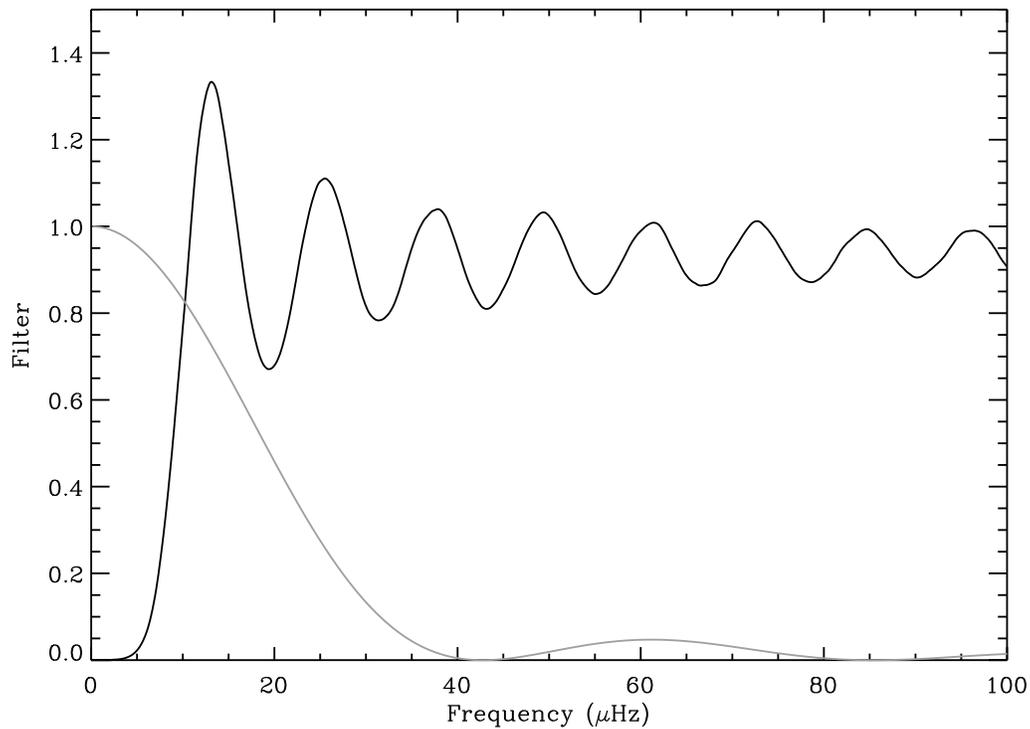}

 \caption{Filter response due to: 2-day Savitsky-Golay filter (black
 line); averaging of data from 29.4-min to 6.5-hr cadence (grey
 line).  A frequency of 11.6 $\mu$Hz corresponds to 1 day.}

 \label{fig:filter}
\end{figure}

Frequency spectra were constructed by adding the PSD due to activity
and granulation. Spectra were computed at the natural frequency
resolution of each of the 3-month-long Quarters of \textit{Kepler}
observations (i.e., $\simeq 0.13\,\rm \mu Hz$), up to the Nyquist
frequency for 29.4-min LC observations.  They were then multiplied by
two functions to allow for the band-pass filtering inherent in
construction of the CDPP. The first function described the high-pass
filtering given by application of the 2-day Savitsky-Golay filter. Its
filter response is plotted in Fig.~\ref{fig:filter} as the black line.
The second function (sinc-squared) described the low-pass filtering
given by averaging of data into 6.5\,hr blocks, i.e., ${\rm sinc}(\pi
\nu T_{\rm av})^2$, with $T_{\rm av} = 6.5\,\rm hr$. Its filter
response is plotted in Fig.~\ref{fig:filter} as the grey line.

The variance (CDPP squared) was then estimated from the sum of the PSD
in the frequency spectrum, multiplied by the natural frequency
resolution (i.e., by invoking Parseval's Theorem
which demonstrates that the total power in
a signal is the same if computed in the time domain or in the Fourier
transform frequency domain, see Press et al. 1992).  The normalization
between the time and frequency domains was double-checked using
artificial time-series data.

The TRILEGAL simulations provide independent stellar parameters
for the $\approx$50\% of cases it returns as physical binaries.
We assume that all such binaries will be unresolved by {\em Kepler}
and treat each binary as one star for which to report noise.
The individual {\em rms} fluctuations for each component of 
binaries are reduced by the sum of the individual fluxes, and 
then recombined in variance space to provide the final noise level.

As argued in \S 3.6 the real observations suffer from small increases
of noise due to superposition of sources which are themselves
variable.  We allow for this relatively minor correction by adopting
the contamination factors used for Fig.~\ref{fig:gm.simback}.  An
underlying probability distribution for the noise multiplication
factors (due to both the known contamination and unknown background
stars) is formed, then for each star in the simulation a random number
is generated and used to fix the multipliers (i.e., to map the random
numbers onto the underlying distribution).
Thus some 25\% of stars have noise levels boosted by 1\%, with about
80\% of stars having noise levels increased by less than 20\%.

\subsubsection{Description of activity}
\label{sec:estact}

We assume that the RMS amplitude of the activity signal, $\sigma_{\rm
act}$, may be estimated from predictions of the Ca~H and K emission
index, $R'_{\rm HK}$. We adopt empirical scaling relations in the
literature -- which use as input the fundamental properties of the
synthetic stars -- to estimate the rotation period, $P_{\rm rot}$, and
the convective turnover time, $\tau_{\rm con}$, at the base of the
convective envelope, from which the
$R'_{\rm HK}$ are predicted from the Rossby number $Ro = P_{\rm
rot}/\tau_{\rm con}$.

We estimate the $P_{\rm rot}$ of the stars using the empirical
relationship of \citet{aig04}, which was derived from analysis of
photometric data taken on stars in the Hyades
(\citealp{radick87,radick95}). The relationship uses the $B-V$ color
and the age, $T$, of each star as input, and assumes a $t^{0.5}$
spin-down in the periods \citep{sku72}. We also tested the empirical
relationship of \citet{car07}, which takes the mass $M$ and age $T$,
with the spin-down calibrated to go like $t^{0.45}$. The convective
turnover time, $\tau_{\rm con}$, was estimated from the empirical
relationship of \citet{noyes84}, using the $B-V$ color as
input. \citet{mamajek08} discuss how this relationship provides a
reasonable match to recent computational predictions based on either
mixing-length theory or full turbulence spectrum treatments of
convection. To go from $Ro = P_{\rm rot}/\tau_{\rm con}$ to $R'_{\rm
HK}$ we used the empirical relationship of \citet{noyes84}. We also
tested the empirical relationship of \citet{mamajek08}.

The final step takes us from $R'_{\rm HK}$ to $\sigma_{\rm act}$. Data
on RMS variations observed in solar-type stars and their relation to
the observed $R'_{\rm HK}$ are presented in \citet{radick98},
\citet{lock07} and~\citet{hall09}.  These data suggest a power-law
dependence of the form $\sigma_{\rm act} \propto (R'_{\rm
HK})^{1.75}$, with the observed RMS variation for the Sun being about
500\,ppm.
As will be noted further below, the ground-based data generally
pertain to timescales of weeks, while we are considering timescales of
several hours.

To check the calibration, we analyzed around 12\,yr of photometric
``Sun-as-a-star'' observations (used also in \S 3.3) made in the green wavelength channel of
the VIRGO/SPM instrument onboard the ESA/NASA SOHO spacecraft
(Fr\"ohlich et al. 1997). The stellar observations referred to above
were made in Str\"omgren $b$ and $y$, and should have a similar
response to the narrowband green SPM data. We divided the 12-yr SPM
lightcurve into contiguous 3-month datasets -- corresponding to the
Quarter-long \textit{Kepler} datasets -- and then reduced the cadence
in each dataset from 60\,sec to 30\,min (by averaging into 30-datum
blocks). The RMS variation due to stellar activity was then estimated
(the estimated contribution from granulation having first been
removed; see Section~\ref{sec:estgran} below). The estimated
$\sigma_{\rm act,\,\odot}$ varied from about 150\,ppm at solar
activity minimum to just over 600\,ppm at solar activity maximum, with
an average value of around 350\,ppm. This average differs from the
500\,ppm solar value returned by the
ground-based stellar observations, which may reflect the impact of
different averaging lengths for the data (a point we return to below,
just after Equation~\ref{eq:sigma1}.  We
therefore apply a multiplicative correction of $(350/500) = 0.7$ to
the empirical scalings suggested by the stellar data.

Two further corrections are applied. First, we make a multiplicative
correction of $\simeq 0.8$ to allow for the fact the \textit{Kepler}
bandpass has significant width in wavelength resulting in a redder
mean response by 20\% compared to the narrowband green SPM channel
(see Section~\ref{sec:intrinsol} above).

Second, we correct for the impact of the angles of inclination
offered by the stars. As discussed in \citet{knaack01}, as the angle
of inclination is decreased from 90\,degrees to the most probable case
of 57\,degrees so variations due to photometry will increase and those
due to Ca~H and K emission will decrease. \citet{knaack01} modelled
the expected changes for a Sun-like star, and predicted that the
aforementioned decrease in angle would lead to a $\approx 5\,\%$
increase in irradiance variability (the \textit{Kepler} wavelength
response is much closer to the irradiance response than it is to the
Str\"omgren $b$ and $y$ response\footnote{As already noted in the
text, a multiplicative factor of ~0.8 corrects for the \textit{Kepler}
response being redder than the VIRGO/SPM green or Stromgren b+y bands.
Analysis of total solar irradiance data collected by the PMO6
instrument on VIRGO/SOHO gives values for RMS variability that are
also about 0.8-times the VIRGO/SPM green-band variability.}). Knaack
et al. predict a negligible decrease in the absolute Ca~H and K flux
($\approx 1\,\%$, a change which we choose to ignore here). The solar
data come from observations where the angle of inclination offered by
the Sun is close to 90\,degrees. To correct the calibration to the
most probable angle of 57\,degrees, we therefore multiply the average
$\sigma_{\rm act,\,\odot}$ inferred from analysis of the solar data by
the factor 1.05.

The stellar data presented in \citet{hall09} follow:
 \begin{equation}
 \sigma_{\rm act} \simeq 10^{(11.5 + 1.75\,\log_{10}
 R'_{\rm HK})}\,\rm ppm.
 \label{eq:sigma0}
 \end{equation}
Application of the multiplicative corrections (combined factor of
$\simeq 0.6$) gives our adopted empirical relation:
 \begin{equation}
 \sigma_{\rm act} \simeq 6 \times 10^{(10.5 + 1.75\,\log_{10}
 R'_{\rm HK})}\,\rm ppm.
 \label{eq:sigma1}
 \end{equation}
Implicit in our adoption of the above is that the exponent of the
power-law dependence (here, 1.75) does not change when the data are
averaged on different timescales,
i.e., the temporal coverage of the ground-based stellar observations
is such that their data provide variability information on timescales
of several months to years, while (in the absence of any other stellar
reference) we have used these data to calibrate changes on much
shorter timescales.  We have not included any
perturbations to Equation~\ref{eq:sigma1} to allow for the fact that
we will be catching different stars in different phases of their
stellar cycles, assuming instead that the impact of this variability
will tend to average out over the ensemble.

With regards to the timescale of the activity, $\tau_{\rm act}$, we
make the gross approximation of applying the solar timescale to all
synthetic stars. Analysis of the green SPM channel data returned an
estimate of $\tau_{\rm act,\,\odot} \simeq 8\,\rm days$, with the
observed PSD showing a reasonable match to the Lorentzian form of
Equation~\ref{eq:psd}. This timescale of course differs from the solar
rotation period of $\simeq 26\,\rm days$, and reflects the complicated
interplay between rotational modulation and the lifetimes of active
regions on a timescale (3\,months) that captures only three rotation
periods. \citet{aig04} suggested that the rotational timescale may
become dominant in more rapid rotators. We therefore tested the impact
of setting $\tau_{\rm act}= P_{\rm rot}$ when $P_{\rm rot} \la 8\,\rm
days$, but found that this did not have a significant impact on the
distribution of predicted stellar noise values.

\subsubsection{Description of granulation}
\label{sec:estgran}

We follow \citet{huber09} and~\citet{cha11b} in assuming that the
timescale for the granulation, $\tau_{\rm gran}$, scales inversely
with the frequency of maximum oscillations power, $\rm \nu_{\rm max}$,
as proposed by \citet{kjeldsen11}. To fix the RMS of the granulation,
$\sigma_{\rm gran}$, we follow the procedure outlined in
\citet{cha11b}. This procedure assumes that the typical size of a
convective granule is proportional to the scale height for an
isothermal atmosphere, and that all granules behave in a statistically
independent manner so that the total RMS fluctuation scales inversely
as the square root of the number of observed granules (e.g.,
\citealp{schwarzschild75,kjeldsen11}). This leads to $\sigma_{\rm
gran} \propto \nu_{\rm max}^{-0.5}$.

In order to fix solar values for $\tau_{\rm gran}$ and $\sigma_{\rm
gran}$, we again analyzed the green SPM channel Sun-as-a-star
data. Fits to the power spectral density in the frequency region below
the solar p modes where the granulation dominates gave average values
of $\tau_{\rm gran,\,\odot} = 220\,\rm sec$ and $\sigma_{\rm
gran,\,\odot} = 75\,\rm ppm$. Since $\nu_{\rm max} \propto
MR^{-2}T_{\rm eff}^{-0.5}$ (\citealp{kjeldsen95,cha08}), the scaling
relations we adopt to estimate $\tau_{\rm gran}$ and $\sigma_{\rm
gran}$ are then given by:
 \begin{equation}
 \tau_{\rm gran} = 220 \left( \frac{M}{\rm M_{\odot}} \right)^{-1}
 \left( \frac{R}{\rm R_{\odot}} \right)^{2}
 \left( \frac{T_{\rm eff}}{{\rm T_{eff,\,\odot}}}\right)^{0.5}\,\rm sec,
 \label{eq:grantau}
 \end{equation}
and
 \begin{equation}
 \sigma_{\rm gran} = 75\, \left( \frac{M}{\rm M_{\odot}} \right)^{-0.5}
 \left( \frac{R}{\rm R_{\odot}} \right)
 \left( \frac{T_{\rm eff}}{{\rm T_{eff,\,\odot}}}\right)^{0.25}\,\rm ppm,
 \label{eq:gransig}
 \end{equation}
with ${\rm T_{eff,\,\odot}}=5777\,\rm K$. Use of the
zero-frequency-centered Lorentzian (Equation~\ref{eq:psd}) to describe
the granulation provides a reasonable description of the observed PSD
at frequencies below the range occupied by the p modes. The Lorentzian
description does however appear to fail at higher frequencies (e.g.,
see \citealp{huber09,karoff11}), which is not a cause for concern here
since the band-limited filtering used to construct the CDPP selects out
PSD only at very low frequencies (see Section~\ref{sec:est} above).

\begin{figure}

 \centering
 \includegraphics[width=0.75\textwidth,clip]{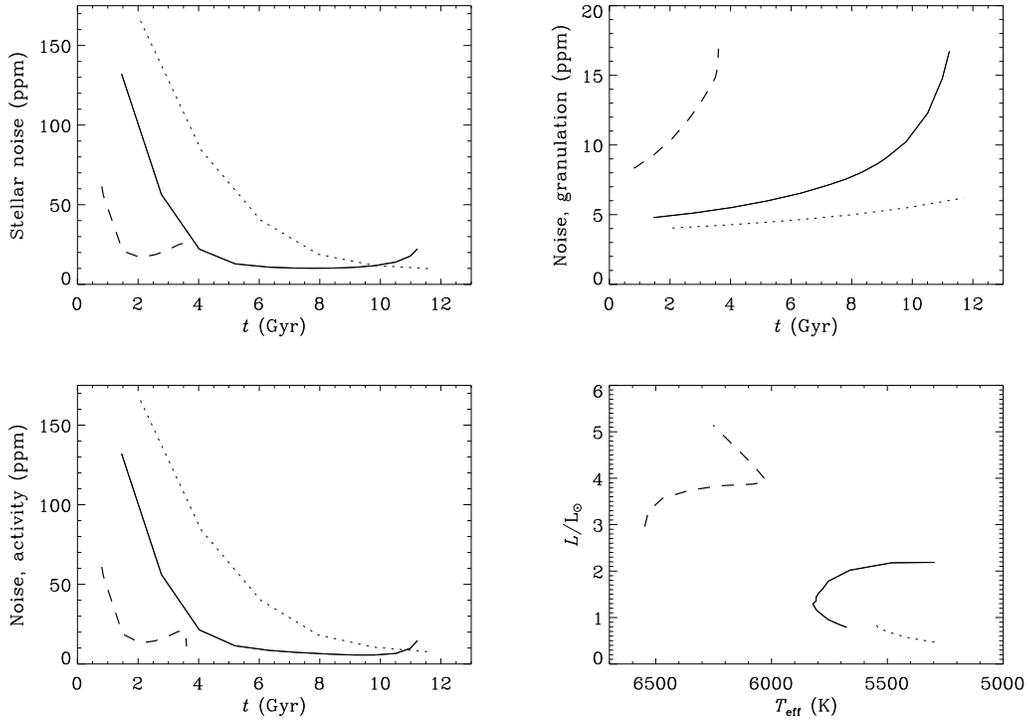}

 \caption{Simulated total stellar noise on 6.5 hour timescales (top left-hand panel) and
 noise due to granulation (top right-hand panel) and activity (bottom
 left-hand panel) versus time, $t$, for three models of mass
 $0.9\,\rm M_{\odot}$ (dotted lines), $1.0\,\rm M_{\odot}$ (solid
 lines) and $1.3\,\rm M_{\odot}$ (dashed lines). The bottom right-hand
 panel shows the models on an HR diagram.}

 \label{fig:evol}
\end{figure}

Fig.~\ref{fig:evol} plots the simulated total stellar noise (top
left-hand panel) and noise due to granulation (top right-hand panel)
and activity (bottom left-hand panel) versus time, $t$, for three
models of mass $0.9\,\rm M_{\odot}$ (dotted lines), $1.0\,\rm
M_{\odot}$ (solid lines) and $1.3\,\rm M_{\odot}$ (dashed lines). The
bottom right-hand panel shows the models on an HR diagram. As the
stars get older the contribution from granulation increases, while
that from activity declines until the stars leave the main sequence.
At this point the shift to lower effective temperature increases the
convective turnover time, $\tau_{\rm con}$, thereby increasing the
Rossby number, $Ro$, thus halting and even reversing the decline of
activity with age \citep{gil85}.  Our exclusion of subgiants from the
stellar sample prevents the latter effect from being significant
here.

 \subsection{Results from synthetic population}
 \label{sec:ressyn}

\begin{figure}

 \centering
 \includegraphics[width=0.75\textwidth,clip]{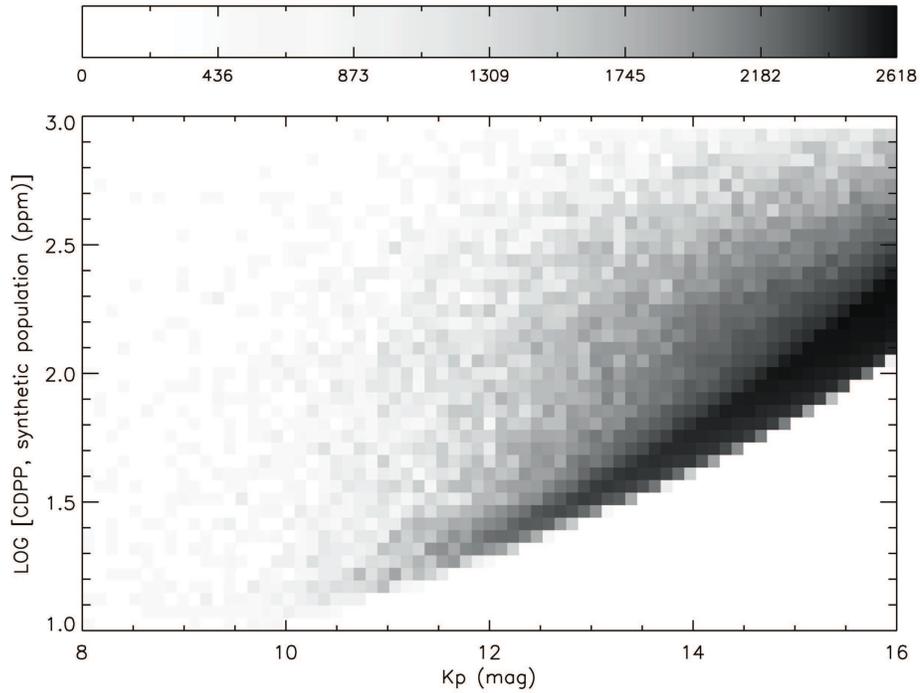}

 \caption{Estimated full CDPP of the synthetic population of
solar-type stars.  The bar at the top defines density of stars
per 0.1 (mag) $\times$ 0.05 (Log CDPP) bin.}

 \label{fig:syn1}
\end{figure}

\begin{figure}

\plottwo{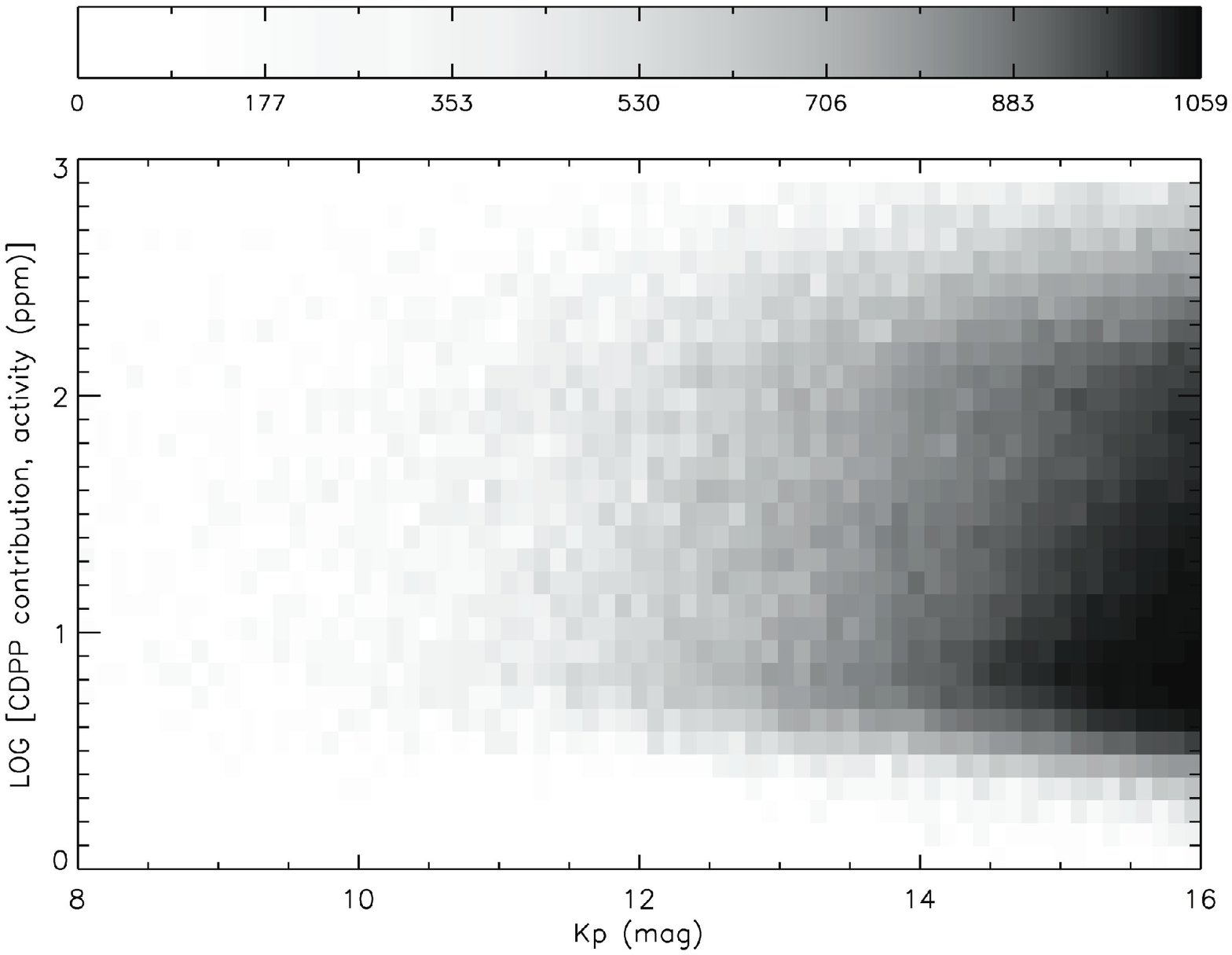}{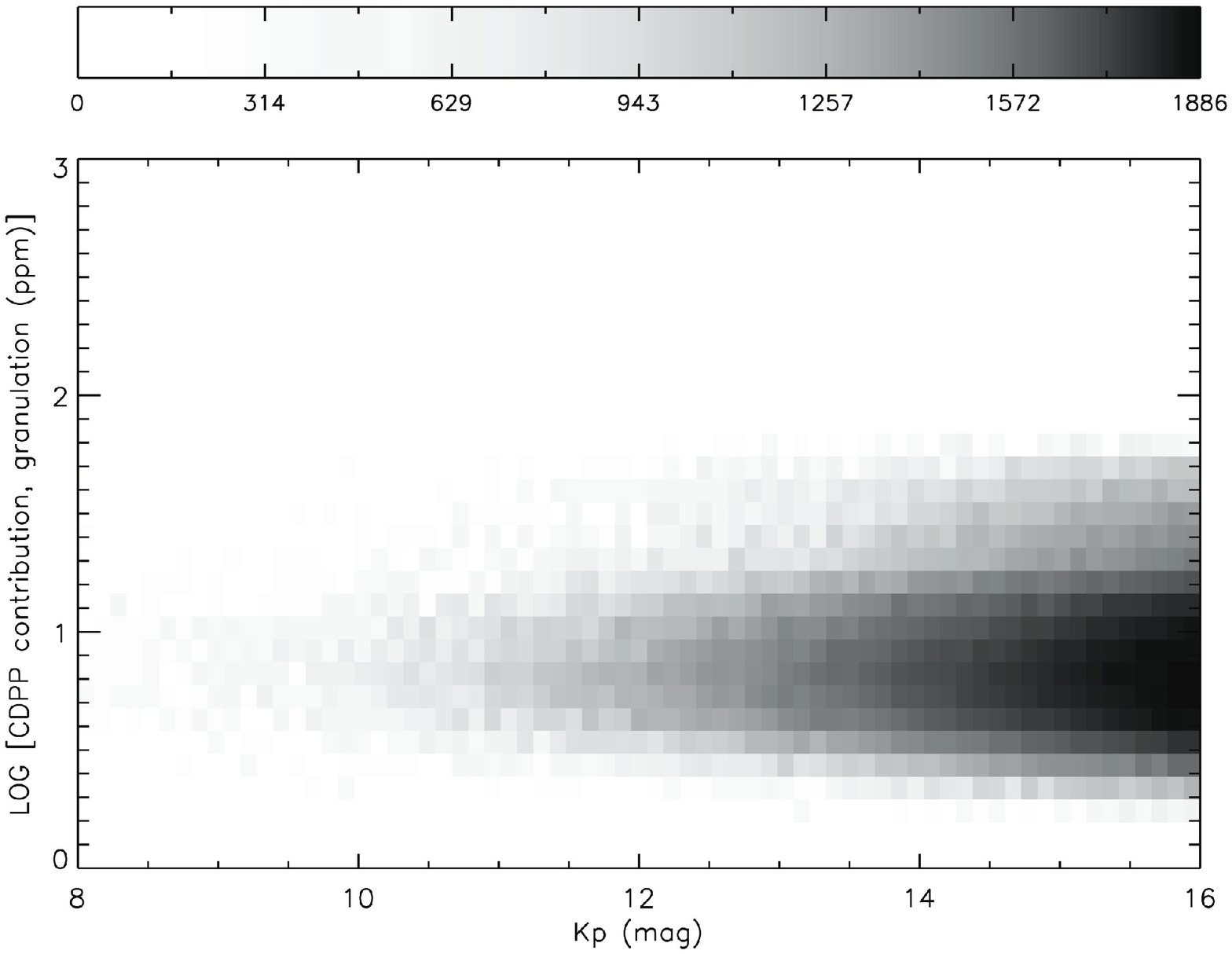}

 \caption{Left-hand panel: Contribution from activity to CDPP of
 solar-type stars in synthetic population. Right-hand panel:
 Contribution from granulation.  Bars at top define the density of
stars per 0.1 (mag) $\times$ 0.1 (Log CDPP) bin.}

 \label{fig:syn2}
\end{figure}

Fig.~\ref{fig:syn1} plots the estimated full CDPP of the solar-type
stars in the synthetic population. To give the full CDPP, we added (in
variance space) a contribution to represent shot noise which was
calibrated on a linear interpolation (random on [0,1] by star) of the
minimal LC noise and upper envelope noise models presented in
\citet{jen10b}.  Fig.~\ref{fig:syn2} shows the respective
contributions from activity (left-hand panel) and granulation
(right-hand panel). The synthetic plot shows the key features present
in the CDPP plot of the real data (Fig.~\ref{fig:gm.cdppo}): a
concentration of stars which map the domain just above the lower-bound
defined by the shot noise limit, and a wider spread of stars at higher
CDPP. The concentration is less well defined in the synthetic plot,
and appears to widen slightly at fainter apparent magnitudes.

The predicted contribution from granulation is plotted in the
right-hand panel of Fig.~\ref{fig:syn2}. The lower swathe of stars
maps the granulation of either single stars or the primary components
of binaries, and is strongly band-limited by the filtering inherent in
construction of the CDPP.
The strong, lower concentration to granularity is evidently in part
responsible for the pile-up of values near 8-12 ppm in the synthetic
stellar noise plot, which suggests a similar contribution for the real
data.  The second swathe
at higher noise maps the extra contribution from the secondary
components of binaries.

\begin{figure}

 \centering
 \includegraphics[width=0.5\textwidth]{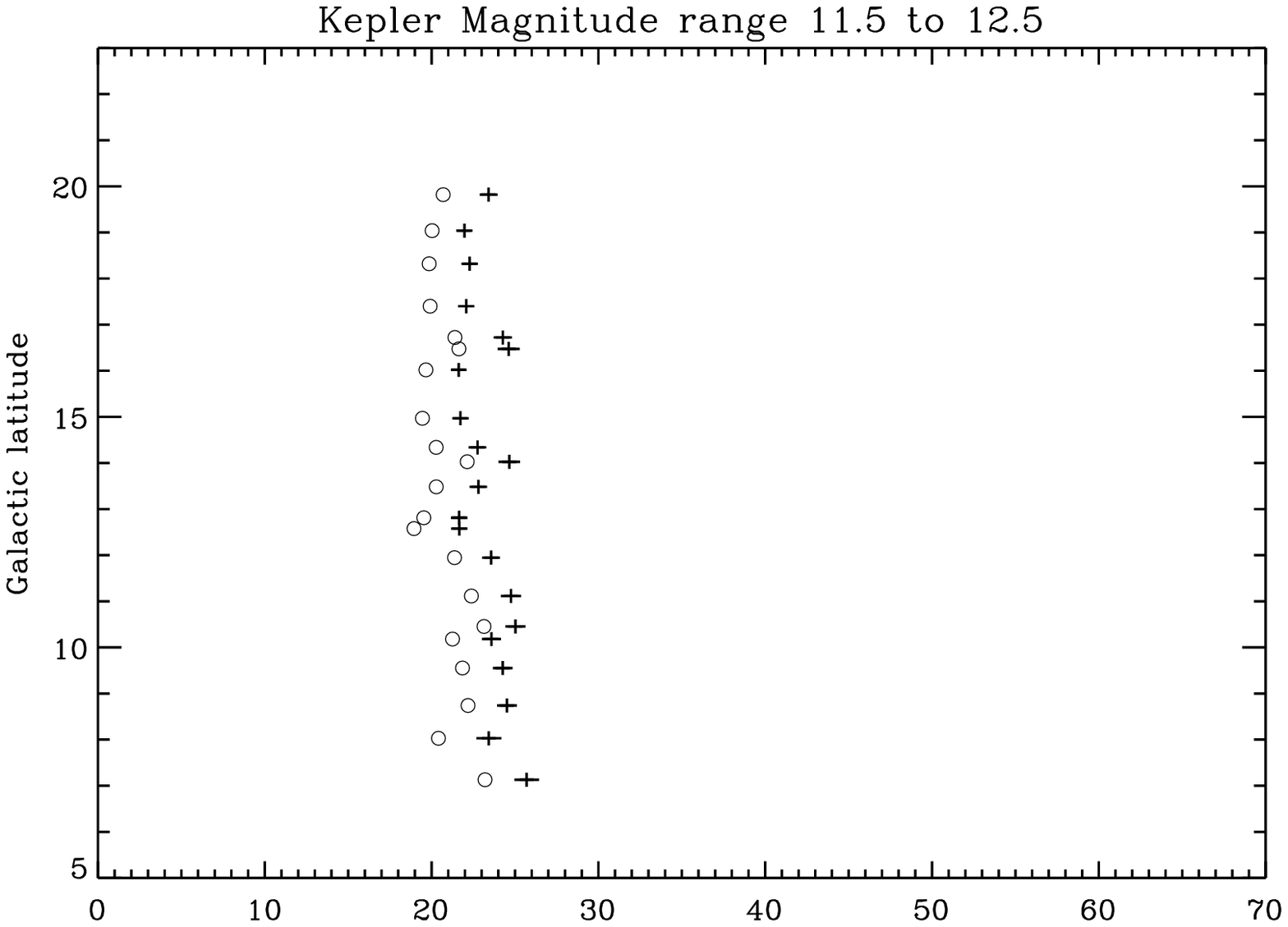}
 \centering
 \includegraphics[width=0.5\textwidth,clip]{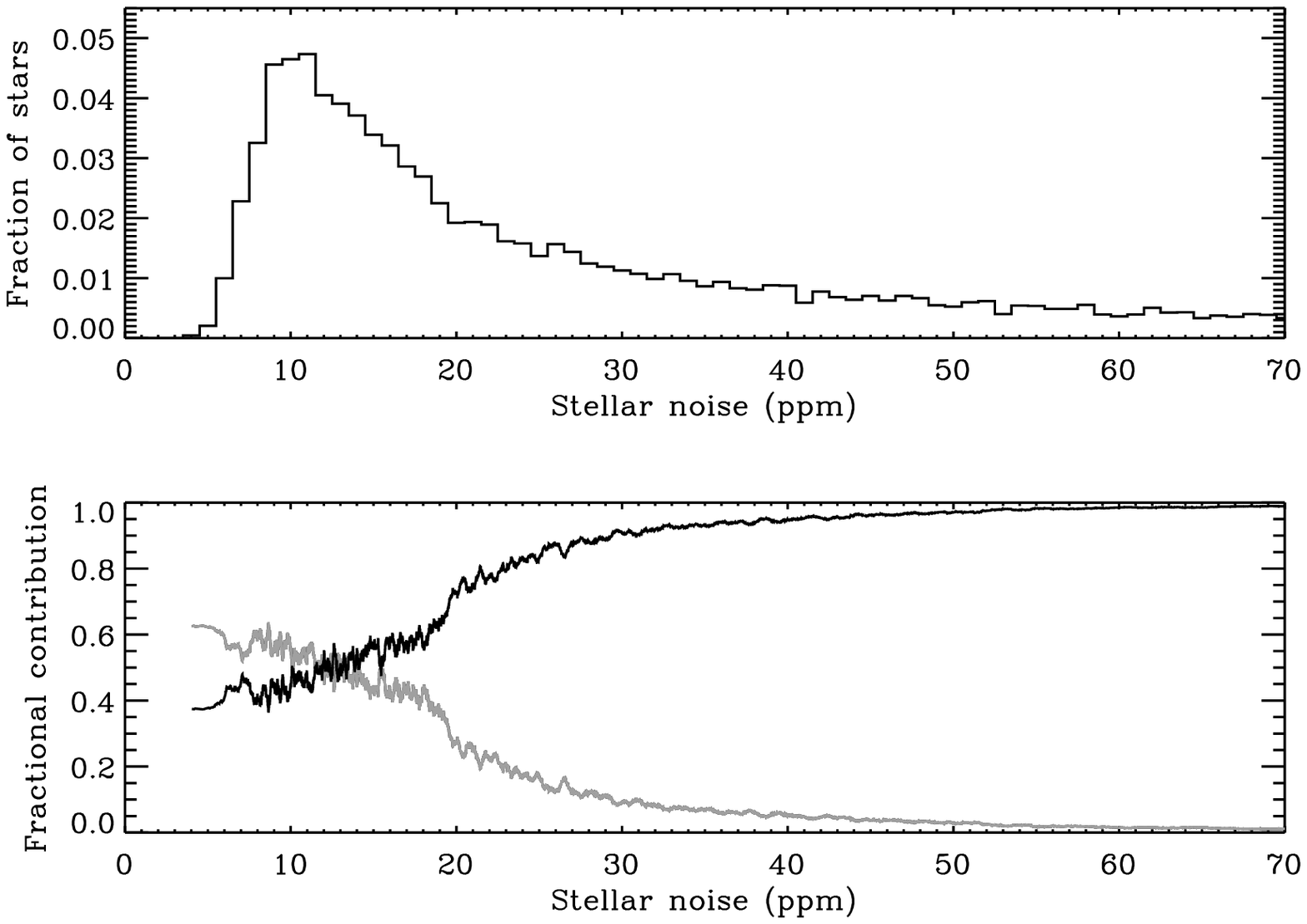}

 \caption{Distribution of stellar noise for solar-type stars in the
synthetic population having \textit{Kepler} apparent magnitudes in the
range 11.5 to 12.5. Top panel: Mean (with error bars) and median (open
circles) stellar noise levels, computed as per the real data in
Fig.~\ref{fig:gm.cdppdist}. Middle panel: Distribution of stellar
noise. Bottom panel: Relative contributions in variance space of
granulation (grey line) and activity (black line).}

 \label{fig:syn3}
\end{figure}

Fig.~\ref{fig:syn3} is analogous to Fig.~\ref{fig:gm.cdppdist}, and
shows the distribution of stellar noise for solar-type stars in the
synthetic population having \textit{Kepler} apparent magnitudes in the
range 11.5 to 12.5. The mean and median stellar noise levels plotted
in the top panel were computed as per the real data.  This is the
result of averaging over 10 independent realizations of TRILEGAL
simulations.

The synthetic stellar noise shows less variation with galactic
latitude, $b$, than does the real data (recall the latter display a
slight decrease in noise with increasing $b$). For intrinsic noise of
30--70 ppm 46\,\% of the artificial stars lie at $b$ $<$ 13 degrees,
while at 17--21 ppm and 7--11 ppm the fraction drops only marginally
to 43\,\%.

The histogram in the middle panel of Fig.~\ref{fig:syn3} bears some
similarity to the histogram of the real data in
Fig.~\ref{fig:gm.cdppdist}, with a common lower envelope and an
extended tail.  The distribution of stars is strongly peaked toward
the lower end of the plotted noise range, although the synthetic
population is peaked at a lower absolute noise than the real
population.  This minor difference in absolute scale is within a
reasonable range of uncertainty for either the simulations, or the
noise decomposition of real data.  We note that the bi-modal signature
seen in the real data is not present when we average results from
independent realizations of the population synthesis, and therefore
the bi-modality is not well established in either the simulated or real data.
Both the simulations and decomposition robustly support
the existence of a broad distribution of intrinsic stellar noise over
$\sim$10 -- 20 ppm near {\em Kp} = 12.  The medians and mean over all
galactic latitudes computed as for the real data in \S 3.5 are 21.6
and 23.7 ppm respectively.  Although more peaked at low values, the
global statistics average some 15\% higher than the observations.

The bottom panel of Fig.~\ref{fig:syn3} shows the relative
contributions in variance space of granulation (grey line) and
activity (dark line).  The respective fractional contributions are
quite similar at low stellar noise; at high noise, it is the activity
which clearly dominates.

As noted previously, we also tested the impact on the predicted
stellar noise of the use of different empirical scaling relations in
estimation of the activity component. Rotation periods predicted with
the relation of \citet{car07} show reasonable agreement with our
adopted relation of \citet{aig04}, although the agreement worsens
progressively for masses above $1.3\,\rm M_{\odot}$ (the \citet{car07}
relation is calibrated for lower masses). When we restrict to
synthetic stars in the range $M < 1.3\,\rm M_{\odot}$, the
distributions of estimated stellar noise are quite similar in
appearance, although the peak in Fig.~\ref{fig:syn3} is shifted to a
slightly lower noise when the \citet{car07} relation is used,
as it gives slightly longer rotation periods (on average).

Our adopted relation to convert from $Ro$ to $R'_{\rm HK}$
\citep{noyes84} and the corresponding relation given in
\citet{mamajek08} show reasonable agreement up to $Ro \simeq 2$,
although differences in the functional forms of the relations -- one
being a cubic function, the other a log-linear function -- lead to
some subtle differences. These differences mean that stars in the peak
shown in Fig.~\ref{fig:syn3} (which was made using the \citet{noyes84}
relation) are re-distributed to slightly higher noise values when the
relation of \citet{mamajek08} is employed, giving the peak a long,
gradually diminishing tail, and making it look less like the real
histogram peak. For $Ro \ga 2$ the relations diverge significantly,
with \citet{mamajek08} giving significantly lower values of $R'_{\rm
HK}$ at a given $Ro$ than \citet{noyes84}. However, it is the
differences at $Ro < 2$ that matter more for the observed
distributions.

\section{SIMULATED AND OBSERVED RESULTS AT {\em Kp} = 14.5}

\begin{figure}

 \centering
 \includegraphics[width=0.5\textwidth]{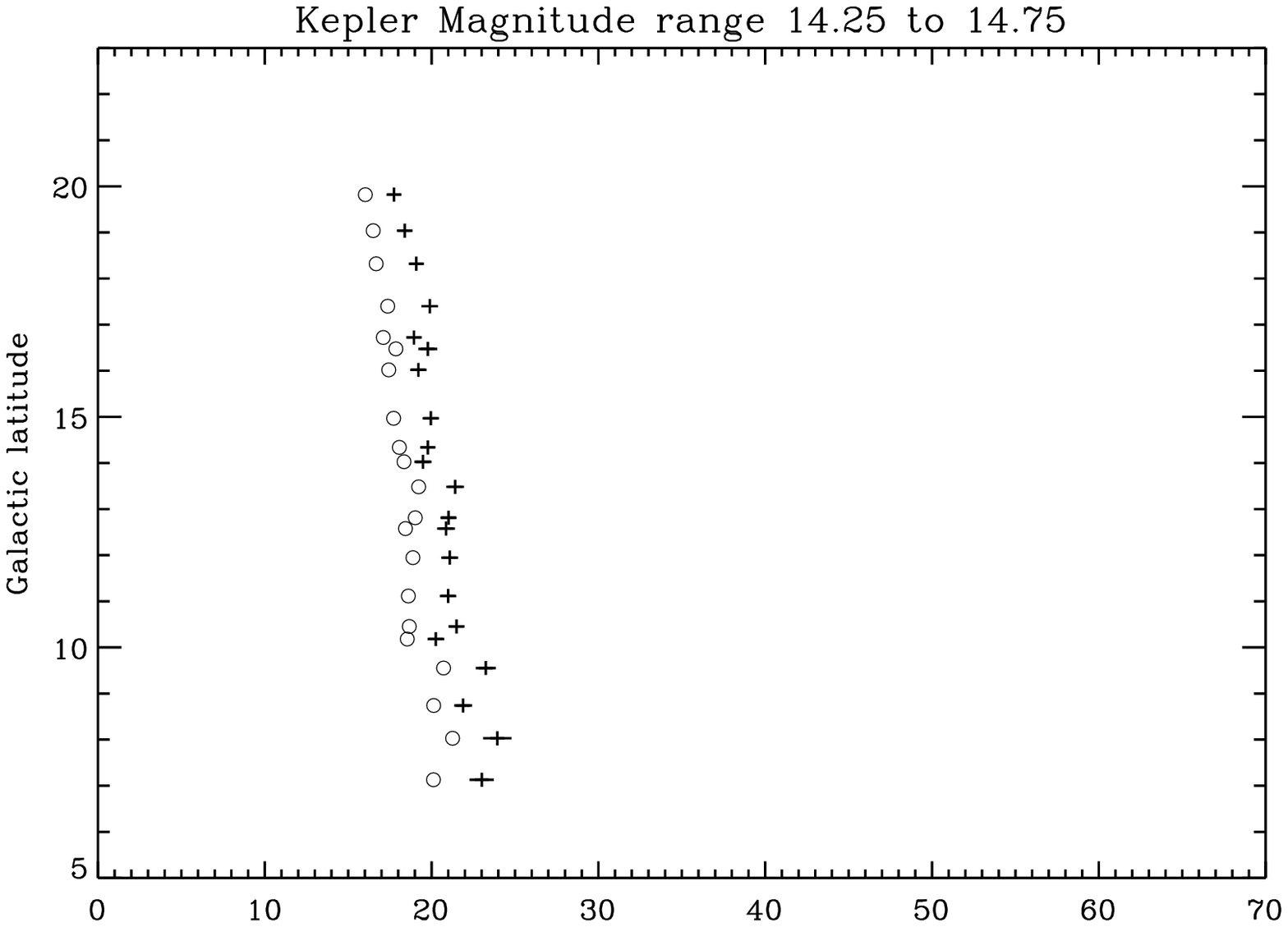}
 \centering
 \includegraphics[width=0.5\textwidth,clip]{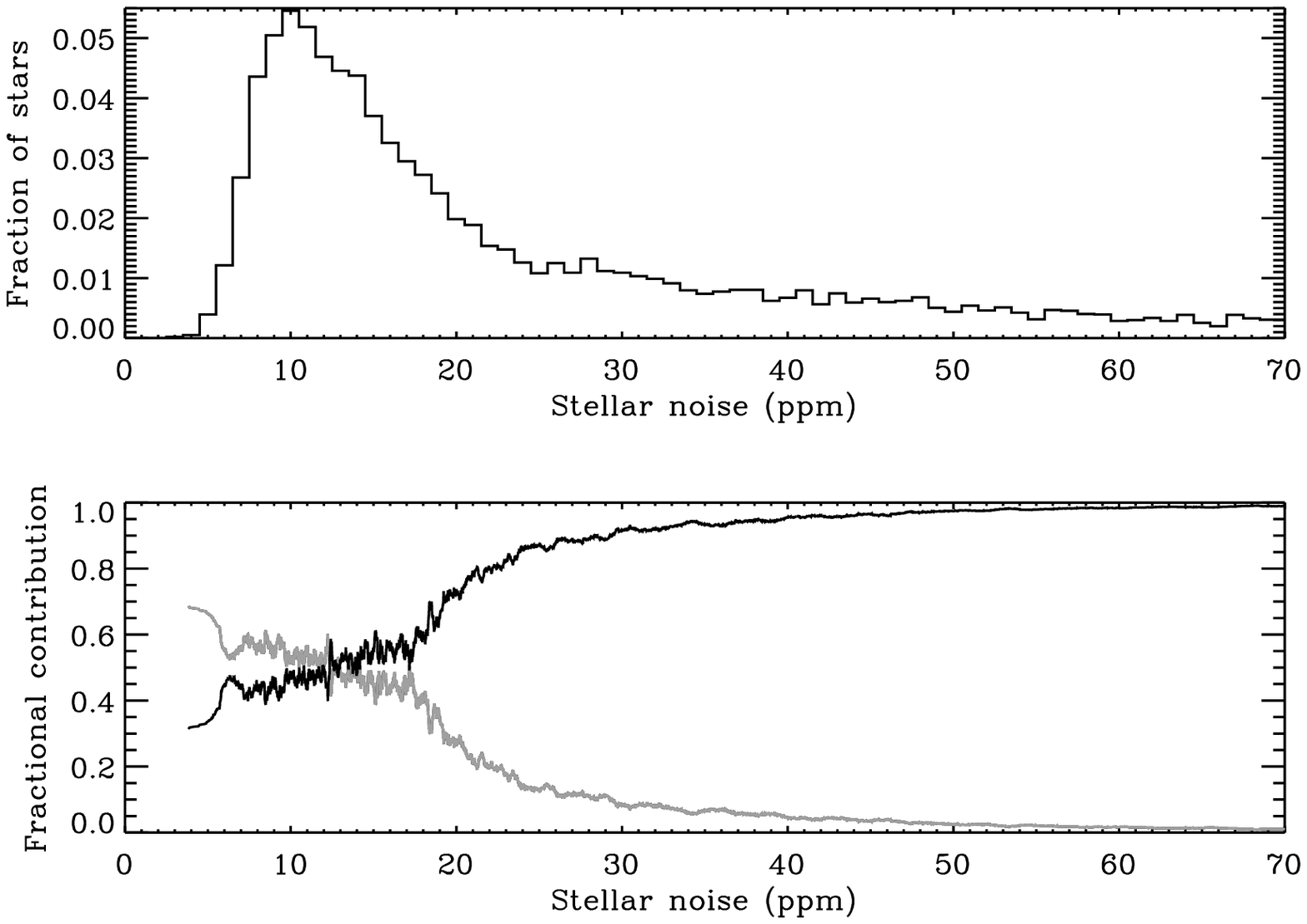}

 \caption{Distribution of stellar noise for solar-type stars in the
synthetic population having \textit{Kepler} apparent magnitudes in the
range 14.25 to 14.75. Top panel: Mean (with error bars) and median
(open circles) stellar noise levels, computed as per the real data in
Fig.~\ref{fig:gm.cdppdist}. Middle panel: Distribution of stellar
noise. Bottom panel: Relative contributions in variance space of
granulation (grey line) and activity (black line).}

 \label{fig:syn14}
\end{figure}

Fig.~\ref{fig:syn14} shows the distribution of stellar noise for the
synthetic population stars in the range $Kp=14.25$ to 14.75. There is
a significant decrease in noise with increasing latitude, $b$, which
follows from clear trends in age, and hence the rotation periods, of
the synthetic stars.  From low to high latitude, the mean age
increases monotonically from 4.5 to 5.5\,Gyr, while the mean rotation
period increases from 13 to 15\,days. For intrinsic noise of 30--70
ppm 52\,\% of the artificial stars lie at $b$ $<$ 13 degrees, while at
17--21 ppm and 7--11 ppm the fraction drops to 47\,\% and 42\,\%,
respectively. These clear trends are much more pronounced than those
in the synthetic population results for $Kp=11.5$ to 12.5, being much
more similar to the observed trends as discussed in \S 3.5.  This may
be taken as another sign \citep{cha11a} that the TRILEGAL-supplied
distribution of fundamental stellar parameters for bright stars in the
{\em Kepler} field differs in detail from actual values.

Comparison with the observed results for this $\times$10 fainter
sample than the $Kp$ = 11.5 to 12.5 set considered earlier is best presented
by a different approach.  For the brighter sample, corresponding
to the point design of the {\em Kepler Mission} \citep{koc10},
the intrinsic stellar noise was the largest term, followed by
the Poisson and other terms that in concert remained only 
slightly larger than the stellar noise.  At the fainter sample
the intrinsic stellar noise is similar, but the other
terms grow such that the stellar noise is a minor 
contributor to the overall error budget.  In this case the 
decomposition for intrinsic stellar noise would be much less stable.
Since the decomposition even at 12th magnitude required
finesse of multiple degeneracies, and resulted in small
fractions of non-physical results (e.g. negative stellar
variances), we opt here to only consider a forward
consideration comparing observed and simulated results.

\begin{figure}
\includegraphics[width=10cm]{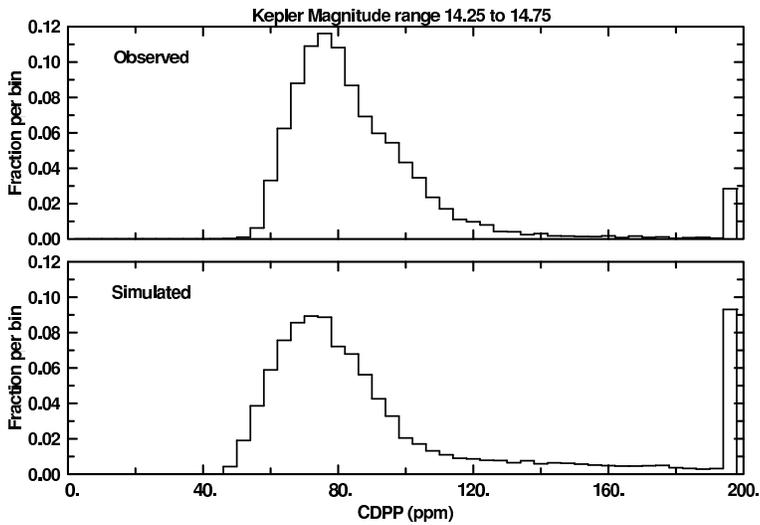}
\caption{Upper panel shows distribution of observed CDPP 
for the {\em Kp} = 14.25 -- 14.75 sample.  The lower panel shows
the distribution of simulated CDPP for the same magnitude
range drawn from Fig.~\ref{fig:syn1}.
\label{fig:gm.14obsim}}
\end{figure}

Fig.~\ref{fig:gm.14obsim} shows the relative distribution of the
as-observed CDPP, contrasted to the distribution of simulated CDPP as
pulled from Fig.~\ref{fig:syn1}.  The observed values and simulations
are in reasonable agreement with a very similar mode, differing only
in distribution details.  Evaluated from values up to 300 ppm the
observed and simulated CDPPs at {\em Kp} = 14.5 have medians of 80.7 and
78.9 ppm respectively.  With means evaluated for stars up to 3$\times$
the median the results are 85.4 and 88.8 ppm for the observed and
simulated stars, respectively.

\section{PROSPECTS FOR CONFIRMATION}

We have argued that much of the observed photometric variation in even
the quiet {\em Kepler} stars arises from intrinsic variability through
granulation and stellar activity on the time scales relevant here.
Both the decomposition approach of \S 3, and the simulation approach
of \S 4 suggest a distribution function for stellar noise
characterized by a broad (possibly bi-modal) peak at low intrinsic
variability, and an extended tail to larger variations.  Since we have
contributed star-by-star measurements for the intrinsic variability it
is possible to provide arguments for follow-up observations that could
test these conclusions.

\subsection{Derivation of Stellar Rotation Period from {\em Kepler} Data}

Fig.~\ref{fig:gm.3exam} suggested that qualitative differences may
exist for the light curves coming from different portions of the
intrinsic noise populations.  In particular light curves for stars
selected based on noise from the extended tail as shown in
Fig.~\ref{fig:gm.cdppdist} should be amenable to derivation of stellar
rotation periods from the {\em Kepler} light curves, e.g. from the
approach of \citet{mos09} applied for CoRoT data.  The same set of
stars could for consistency have $v$sin$i$ measured with high
resolution spectroscopy.  Confirmation that this set of stars is more
rapidly rotating than the Sun would provide support for the primary
assumption of this paper, that the noise for these stars is typically
intrinsic, and that it arises from high activity associated with
stellar youth and still rapid rotation.

\subsection{Spectroscopic Measurement of Activity Indices}

A primary component of the simulations of \S 4 involves model
predictions of the $R'_{\rm HK}$ Ca~H and K emission index.  Measures
for this could be obtained for the three ranges of intrinsic
variability as in Fig.~\ref{fig:gm.cdppdist} by observing, at high
spectroscopic resolution, several stars from the 7--12, 14--19, and
30--50 ppm ranges.  For these corresponding sets in
Fig.~\ref{fig:syn3} the mean $\log R'_{\rm HK}$ values are: $-5.3$,
$-5.2$, and $-4.9$ respectively.  Mean ages are, respectively, 6.6,
4.7 and 2.8\,Gyr, while mean rotation periods are 13.9, 12.7 and
12.3\,days.
Our decompositions and simulations predict that stars drawn from these
subsets will have progressively increasing levels of magnetic activity
that serve as causal agents for the inferred photometric variability.
The fraction of stars from Fig.~\ref{fig:syn3} as quiet as the 
Sun ($\leq$ 12.5 ppm) is 23\%, which contrasts with expectations
\citep{bas05} that two-thirds of solar type stars would be as 
quiet as the Sun.  The TRILEGAL simulations have only 24\% of the
solar-type stars older than 4.6 Gyr, and 52\% are significantly
younger than the Sun at $<$ 3 Gyr.
The fraction of observed stars in Fig.~\ref{fig:gm.cdppdist}
below 12.5 ppm is also 23\%.
We note, however, that the simulations both at {\em Kp} = 12 and 14.5
have larger extended noise tails than the observations -- 
consistent with the age distribution in the simulations
being too young.

\subsection{Asteroseismic Age Determinations}

Perhaps the highest fidelity test to confirm the basic
premises of this paper could be obtained by using 
{\em Kepler} short cadence data for a brighter subset
of stars in these different intrinsic noise ranges to
define stellar ages.  Recent asteroseismic applications for 
the transiting exoplanet host HAT-P-7 \citep{chrd10}
with {\em Kepler} provided an age estimate of 2.14 $\pm$ 0.26 Gyr,
while similar results using {\em HST} for HD 17156 
\citep{gil11} gave 3.2 $\pm$ 0.3 Gyr.  \citet{met10}
have provided a similarly accurate age estimate for
KIC 11026764 near 6 Gyr good to 15\%, with prospects
in slightly evolved cases like this showing mixed modes
for verifying the age even better.  While it is unlikely
that high fidelity age estimates will follow from the hundreds
of dwarf stars currently having oscillation detections
with {\em Kepler} \citep{cha11a}, it would be reasonable
based on current results to expect such results for
about 100 targets.  Obtaining these age determinations is a primary
goal of the {\em Kepler} asteroseismology program, independent
of the minor role having such would play in testing the
premises in this paper.

\section{SUMMARY}

We have shown that the noise levels resulting for 
{\em Kepler} observations can be decomposed into a 
few terms:  basics such as Poisson statistics and readout noise,
an instrument term that depends on channel number over the 
84 amplifiers, a temporal term that depends on specific 
observing conditions encountered during individual Quarters
of observation, and intrinsic stellar noise.
The dominant term for roughly solar-type 12th magnitude 
stars in the overall noise budget is found to follow from
the stars themselves.  Excess instrument noise does 
exist, but is more-or-less in line with expectations, and
contributes little to the overall noise within which 
{\em Kepler} planet searches must be conducted.
By contrast the intrinsic stellar noise, although still
very modest at less than 20 ppm for the highest concentration of stars,
is a factor of two larger than had been budgeted for.
This results in CDPP estimates for {\em Kp} $\sim$ 12 that are 50\%
larger than anticipated.

We have shown via simulations of expected fundamental 
stellar parameters over the {\em Kepler} field of view,
followed by projections of stellar rotation and resulting
activity levels, that stellar variability consistent 
with that observed can be reproduced to first order.
These simulations at {\em Kp} $\sim$12 produce a broad distribution
of stars with intrinsic noise levels over $\sim$10 -- 20 ppm that
is consistent with that derived directly from the {\em Kepler} data.

CoRoT provided fundamental advances in time-series photometry
securely establishing that most, if not all red giants are 
variable and reaching impressive new precision levels for 12th
magnitude dwarfs \citep{aig09} sufficient to allow searches for 
small planet transits.
{\em Kepler} has taken this a large step further and is the first mission capable of quantifying
the variability of large numbers of stars to the small
levels by which the Sun is known to vary.
{\em Kepler} will continue to provide exciting new insights into
the astrophysics of quiet stars, and their galactic distributions.
While we are not
surprised to have learned new things from this new
observational capability, the fact that the stars are
more variable than expected has a significant influence
on the ability to readily detect Earth-analog planet 
transits where the expected signal per transit is only
a few times the inferred noise level on comparable 
time scales.
Observing for a longer time baseline can compensate for the 
loss of transit detection sensitivity from the higher than
anticipated stellar noise.





\acknowledgments

\medskip

{\em Kepler} was competitively selected as the tenth Discovery
mission.  Funding for this mission is provided by NASA's Science
Mission Directorate.  RLG has been partially supported by NASA
co-operative agreement: NNX09AG09A.  WJC, YE and AM acknowledge
support from the UK Science and Technology Facilities Council (STFC).
A large number of people have contributed to make this Mission a
success, and are gratefully thanked for having done so.  We thank
Jessie Christiansen for providing tabulated values of SOC products.
Georgi Mandushev and Andrej Prsa provided summaries of Besancon model
star counts.  We thank
David Soderblom for discussion of expected stellar variability.
We thank the referee, John Stauffer, for several perceptive remarks
and suggestions which have served to improve the presentation.



{\it Facilities:} \facility{{\em Kepler}}.

\clearpage



\clearpage

\clearpage

\clearpage

\clearpage

\begin{table}
\begin{center}
\caption{Quarter-to-Quarter excess variance.\label{tbl-1}}
\begin{tabular}{lccccc}
\tableline\tableline
Quarter & Q2 & Q3 & Q4 & Q5 & Q6 \\
\tableline
Variance & 210.46 & 105.82 & 44.52 & 0.00 & 29.89 \\
Noise & 14.51 & 10.29 & 6.67 & 0.00 & 5.47 \\
\tableline
\end{tabular}
\tablecomments{Variances in ppm$^2$ and noise levels in ppm over
the five Quarters of {\em Kepler} data analyzed.}
\end{center}
\end{table}

\begin{table}
\vspace*{-0.1in}
\begin{center}
\caption{Channel-to-channel variance and stellar ensemble offsets.\label{tbl-2}}
\begin{tabular}{rrrrrrrrrrr}
\tableline\tableline
Quartet & Ch1 & Ch2 & Ch3 & Ch4 & Var1 & Var2 & Var3 & Var4 & Stellar & N$_{star}$ \\
\tableline
 1 & 1 & 29 & 53 & 81 & 140.90 & 54.99 & 192.28 &  0.00 & 402.58 &  110 \\
 2 & 2 & 30 & 54 & 82 &  98.56 & 142.10 & 210.43  & 0.00 & 239.58 & 110 \\
 3 & 3 & 31 & 55 & 83 & 247.34  & 54.76 & 227.10  & 0.00 & 388.34 & 109 \\
 4 & 4 & 32 & 56 & 84 & 216.59 & 209.20 & 250.18  & 0.00 & 371.60 & 106 \\
 5 & 5 & 33 & 49 & 77 &  0.00 &  89.25 &  67.81 &  34.47 & 255.40 & 137 \\
 6 & 6 & 34 & 50 & 78 & 30.65  & 30.40  & 11.43 &  0.00 & 375.71 & 120 \\
 7 & 7 & 35 & 51 & 79 & 222.62 &  27.85 &  0.00 & 122.56 & 415.36 &  99 \\
 8 & 8 & 36 & 52 & 80 & 99.47 & 175.74 & 107.77 &  0.00 & 459.71 & 116 \\
 9 & 9 & 13 & 69 & 73 &  0.00 &  48.47 &  8.15  & 48.41 & 341.25 & 121 \\
10 & 10 & 14 & 70 & 74 & 107.84 &  21.19 &  0.00  & 69.57 & 446.70 & 114 \\
11 & 11 & 15 & 71 & 75 & 158.94 &  42.07 &  0.00  & 131.08 & 497.34  & 109 \\
12 & 12 & 16 & 72 & 76 & 32.50  & 96.09 &  38.06 &  0.00 & 485.31 & 109 \\
13 & 17 & 25 & 57 & 65 & 12.23 &  28.59 &  29.61 &  0.00 & 286.63 & 134 \\
14 & 18 & 26 & 58 & 66 & 21.57  & 74.67 & 194.93 &  0.00 & 318.53 & 114 \\
15 & 19 & 27 & 59 & 67 & 53.40 &  36.42 &  3.79 &  0.00 & 337.93 & 135 \\
16 & 20 & 28 & 60 & 68  & 0.00 &  59.28 &  56.96 &  39.63 & 346.57 & 113 \\
17 & 21 & 37 & 45 & 61 & 12.29 &  69.33 &  0.00 &  51.35 & 403.00 & 110 \\
18 & 22 & 38 & 46 & 62 & 39.93 &  65.51 &  0.00 & 154.15 & 399.32 & 117 \\
19 & 23 & 39 & 47 & 63  & 0.00 &  65.34 &  10.18 &  28.90 & 402.67 & 133 \\
20 & 24 & 40 & 48 & 64  & 0.00 & 106.90 &  43.89 &  76.42 & 334.28 & 124 \\
21 & 41 & 42 & 43 & 44 &  0.00 &  35.26 &  61.98 & 102.74 & 435.70 & 100 \\
\tableline
\end{tabular}
\tablecomments{After a leading column with serial number,
the next four columns establish channel numbers included,
which the next four columns provide by-channel variance 
excesses.  The final two columns are median variance of 
the stellar ensemble for the quartet and the number of stars.}
\end{center}
\end{table}

\begin{table}
\vspace*{-0.2in}
\begin{center}
\caption{Number of stars per 19.6 {\em Kepler} pixel area.\label{tbl-3}}
\begin{tabular}{crrrrrrrrrr}
\tableline\tableline
{\em Kp} & 13.5 & 14.5 & 15.5 & 16.5 & 17.5 & 18.5 & 19.5 & 20.5 & 21.5 & 22.5 \\
\tableline
$b$ = 19 & 0.008 & 0.016 & 0.03 & 0.04 & 0.06 & 0.09 & 0.12 & 0.15 & 0.19 & 0.21 \\
$b$ = 13 & 0.012 & 0.025 & 0.05 & 0.09 & 0.14 & 0.20 & 0.25 & 0.33 & 0.42 & 0.52 \\
$b$ =  7 & 0.028 & 0.057 & 0.12 & 0.23 & 0.45 & 0.78 & 1.17 & 1.49 & 1.78 & 2.22 \\
\tableline
\end{tabular}
\tablecomments{Besancon model star counts \citep{rob03} for average
number of background blended star per buffered {\em Kepler}
photometric aperture for 12th magnitude stars.}
\end{center}
\end{table}

\begin{table}
\vspace*{-0.2in}
\begin{center}
\caption{Global roll up of noise terms.\label{tbl-4}}
\begin{tabular}{lrrr}
\tableline\tableline
Component & Variance (ppm$^2$) & Noise (ppm) & Baseline Noise (ppm) \\
\tableline
Intrinsic stellar & 380.5 & 19.5 & 10.0 \\
Poisson + readout & 283.0 & 16.8 & 14.1 \\
Intrinsic detector & 116.2 & 10.8 & 10.0 \\
Quarter dependent & 60.1 & 7.8 & --- \\

Total & 839.8 & 29.0 & 20.0 \\
\tableline
\end{tabular}
\tablecomments{The Q2 -- Q6 variance (and derived noise)
contributions for the {\em Kp} = 11.5 -- 12.5 stellar sample
discussed at length in the text.  For the Baseline, design
noise terms, the readout noise is accounted within the Intrinsic
detector term.}
\end{center}
\end{table}

\begin{table}
\vspace*{-0.1in}
\begin{center}
\caption{Percentage of quiet and noisy bright stars.\label{tbl-5}}
\begin{tabular}{lrrrrrrrrr}
\tableline\tableline
Spectral type & A0 & A5 & F0 & F5 & G0 & G5 & K0 & K5 & M0 \\
\tableline
Central T(K) & 10700 & 8250 & 7375 & 6550 & 5925 & 5575 & 5075 & 4325 & 3625 \\
\% at $<$ 10 ppm & 20.5 & 10.9 & 5.9 & 0.7 & 0.0 & 1.2 & 5.4 & 0.0 & 0.0 \\
\% at $>$ 50 ppm & 39.8 & 39.0 & 72.6 & 82.3 & 30.3 & 23.8 & 40.6 & 76.0 & 92.9 \\
No. stars & 88 & 46 & 51 & 305 & 267 & 84 & 74 & 24 & 14 \\
\tableline
\end{tabular}
\tablecomments{Effective temperature divisions between bins are at 8900, 7800,
6950, 6150, 5700, 5450, 4700, and 3950 K, with A0 extending to 12500
K, and M0 to 3300 K. ‘No. stars’ is the total count per spectral bin.}
\end{center}
\end{table}


\end{document}